\documentclass[%
 amsmath,amssymb,
 aps,
prd
]{revtex4-2}

\usepackage{graphicx}% Include figure files
\usepackage{dcolumn}% Align table columns on decimal point
\usepackage{bm}% bold math
\usepackage{hyperref}% add hypertext capabilities
\usepackage{natbib}
\usepackage{mathtools}
\usepackage{physics}
\usepackage{subcaption}
\usepackage{mathrsfs}

\begin{document}

\title{Big-Bounce in Quantum f(R)-Cosmology: Polymer Dynamics with Internal Time}

\author{Maria Luisa Limongi}
 \email{limongi.1744359@studenti.uniroma1.it}
\affiliation{%
    Department of Physics, Sapienza University of Rome, Piazzale Aldo Moro, 2, 00185 Roma, Italy
}%

\author{Simone Lo Franco}
 \email{simone.lofranco@uniroma1.it}
\affiliation{%
    Department of Physics, Sapienza University of Rome, Piazzale Aldo Moro, 2, 00185 Roma, Italy
}%
\affiliation{%
    INFN Section of Rome, Sapienza University of Rome, Piazzale Aldo Moro, 2, 00185 Roma, Italy
}%

\author{Giovanni Montani}
\email{giovanni.montani@enea.it}
\affiliation{
    ENEA, Fusion and Nuclear Safety Department, C.R. Frascati, Via E. Fermi, 45, Frascati, 00044, Roma, Italy
}%
\affiliation{
    Department of Physics, Sapienza University of Rome, Piazzale Aldo Moro, 2, 00185 Roma, Italy
}%

\author{Sebastiano Segreto}
 \email{sebastiano.segreto@uniroma1.it}
\affiliation{%
    Department of Physics, Sapienza University of Rome, Piazzale Aldo Moro, 2, 00185 Roma, Italy
}%

\date{\today}% It is always \today, today,
             %  but any date may be explicitly specified

\begin{abstract}
We construct the Hamiltonian formulation of the isotropic Universe in a generic metric $f(R)$-theory in the Jordan frame. We canonically quantize the Universe volume via a polymer formulation, and we adopt the scalar field naturally arising from this scenario as a physical clock. Being within the limit of cut-off values of the space volume, we are legitimized to neglect, at first approximation level, the self-interacting potential term associated with the scalar field. We first study the semi-classical polymer dynamics, outlining the emergence of a bouncing cosmology, both in the internal as well as in the synchronous time. In this latter time variable, we are also able to compare the obtained picture with that of a standard polymer Big-Bounce. We see that in the studied case, the collapsing and expanding branches are no longer symmetric with respect to the minimum volume configuration. Then, we fully quantize the system dynamics in the momentum representation, constructing a suitable dynamical Hilbert space and setting up the dynamics of localized wave packets. The mean value dynamics, both for the momentum and volume spaces, is characterized by a bouncing dynamics as described via the internal time, which closely resembles that one obtained in Loop Quantum Cosmology and Polymerization, respectively.
\end{abstract}
\maketitle
\section{Introduction}
The consistency and productivity of General Relativity is limited by the existence of solutions of the Einstein equations which contain curvature singularities \cite{Misner:1973prb}. The most important examples of these unphysical divergences of the tidal forces are: the Black Hole singularity \cite{Carter:2009nex} and the Big-Bang in the cosmological setting \cite{Montani:PrimCos,Weinberg:2008zzc}. Moreover, it was possible to demonstrate under which boundary conditions, a physical singularity or a pathology of the geodesic flow can arise \cite{Hawking:1973uf,Montani:PrimCos}, and the Landau School was able to construct the asymptotic behavior of a generic inhomogeneous Universe \cite{BKL_1971,BKL_1982} in the case of the cosmological singularity (see also \cite{kirillov_93,Montani_1995, montani_imponente_01,montani_benini_04, MONTANI_2008, LOFRANCO2025100463}). These shortcomings of General Relativity led to the formulation of the conjecture that new physics is needed in some domains of the gravitational field strength. In particular, two main directions have been followed to solve the singularity problem: i) dealing with possible quantum effects in the gravitational field dynamics \cite{montani_cqg, Thiemann:2007pyv}; ii) altering the gravitational Lagrangian to address extended gravity formulations \cite{Sotiriou_2010,Capozziello:2009nq,Nojiri:2017ncd}. \\ \\%
Quantum formulations in the metric approaches did not get clear evidence for the singularity removal in the behavior of the mean values \cite{Blyth:1975is,kirillov_97,montani_benini2006inhomogeneous}, for a revised scenario see \cite{Giovannetti:2022qje, giovannetti_maione, LoFranco:2024nss}. Very different it has been the story when a connection approach, based on the Ashtekar-Barbero-Immirzi variable \cite{Ashtekar:1986yd,BarberoG:1994eia,Immirzi:1996di}, is considered in the so-called Loop Quantum Cosmology \cite{Ashtekar_2011}. In this formulation, despite being not completely settled down (see for related discussions \cite{Bojowald_2019, Bojowald:2000pk, Cianfrani:2008zv,Cianfrani:2011wg,Bruno:2023aco,Bruno:2023all,Alesci:2013xd,Alesci:2014rra,Alesci:2014uha}), a clear picture for a bouncing cosmology emerged \cite{Bojowald:2002gz,Bojowald:2004af,Ashtekar:2005qt,Ashtekar:2006wn, Ashtekar_2011,Bombacigno:2016siz}, i.e. a configuration in which the mean value of a semi-classical Universe reaches a minimum value, separating a collapsing and expanding branch. In modified gravity theories, the removal of the singularity is not a systematic feature, depending on the specific extension of the Einstein-Hilbert action, as well as on the particular cosmological model being considered. For a discussion of non-singular cosmologies in $f(R)$-theories, see \cite{Bamba:2013fha,Odintsov:2014gea,Odintsov:2015uca,Bamonti:2021jmg}. However, when the torsion field is included in the Lagrangian, i.e., a non-zero antisymmetric affine connection, evidence for a Big-Bounce arises, see for instance \cite{Barragan:2009sq,Barragan:2010qb,Bombacigno:2021bpk}. \\ \\%
The idea that both a modified Einstein-Hilbert Lagrangian and the inclusion of quantum effects are needed to describe extreme regimes of gravity, as well as regulate their singularities, has acquired increasing interest. See, for instance, String Theory approaches \cite{CICOLI20241}, even though the gravitational interaction is only perturbatively quantized in these formulations, while a development of quantum $f(R)$ gravity via Loop Quantum Cosmology was carried on in \cite{Amoros:2014tha}. In \cite{DeAngelis:2021afq}, see also \cite{DeAngelis:2022qhm}, a metric $f(R)$-theory is quantized via a canonical metric formulation for the case of an isotropic Universe. In this formulation, the non-minimally coupled scalar field, emerging in the so-called ``Jordan frame'', has been adopted as internal time for quantum cosmology, with interesting implications on the emergence of a possible classical Universe (see also \cite{Bamonti:2021jmg}). \\ \\%
In this paper, we pursue a similar scenario to that one in \cite{DeAngelis:2021afq}, but we adopt a polymer quantum mechanical formulation for the Universe volume dynamics  \cite{Corichi:2007tf,Barca:2021qdn} (see also \cite{Montani:2018uay,Giovannetti:2020nte} for the discussion of the polymerized isotropic Universe in Ashtekar-Marbero-Immirzi variables and metric variables, respectively). This study aims to investigate whether a bouncing cosmology naturally emerges when the non-minimally coupled scalar field to the Universe volume is taken as a physical clock. Furthermore, we intend to investigate the possible similarities and differences to a Big-Bounce, as described in Einsteinian polymer quantum cosmology. \\ \\%
The paper is structured as follows: i) First, we present the theoretical framework used in this work by making a brief introduction on $f(R)$ modified gravity theories and on Polymer Quantum Mechanics; ii) In the following section, we exploit these formalisms to compute the polymer semiclassical evolution of the homogeneous and isotropic Universe; iii) Then, we move to the quantum mechanical picture. Relying on the ADM reduction procedure, we introduce the reduced quantum Hamiltonian of the system. We then construct localized wavepackets and compute the time evolution of volume and approximated momentum mean values. The results are then compared with what we obtained in the semiclassical discussion; iv) In the last section, we summarize and analyze the results presented in this work.  
\section{Theoretical framework}
In this section, we present the theoretical framework in which the dynamics of the homogeneous and isotropic Universe (FLRW model) is studied. The cornerstone of this treatment is the Arnowitt-Deser-Misner (ADM) formalism, where the space-time manifold is described by a $1+3$ geometry, i.e., $\mathcal{M}=\mathbb{R} \times \Sigma$ with $\Sigma$ being the three-dimensional spatial manifold. In this formalism, the line element of the closed FLRW model takes the form
\begin{equation}
    ds^2=N^2(t)\, dt^2 - a^2(t) \left[ \frac{dr^2}{1-Kr^2} +r^2\left(d\theta^2 + sin^2\theta\, d\varphi^2 \right) \right]\,,
\end{equation}
where $a(t)$ is the isotropic scale factor, $N(t)$ is the {\it lapse function}, $K$ is the spatial curvature, differentiatng between the three different kinds of homogeneous and isotropic models depending on its sign, and $\{r,\theta, \varphi\}$ is the usual set of spherical coordinates. In this framework, the choice of the reference system corresponds to the choice of the lapse function $N(t)$. The ADM splitting of the metric opens the possibility for the definition of the Hamiltonian formulation of GR, where gravity is described by a first-class constrained Hamiltonian system. When applied to homogeneous geometries, like the one under investigation in this work, the dynamics of the system involves only a finite number of geometrical degrees of freedom (composing the so-called {\it mini-superspace}), and we are left with only one non-trivial constraint, which is the super-Hamiltonian constraint. Moving to the quantum mechanical framework, we apply the Dirac quantization scheme to the Hamiltonian formulation of GR, promoting generalized coordinates to operators acting on the Universe wave function. Since the Hamiltonian of GR is a linear combination of constraints, the Universe wave function is independent of the coordinate time, and the notion of time has to be reintroduced at the quantum level. To avoid this issue, we rely on the ADM reduction of the dynamics, i.e., we identify one of the system's degrees of freedom as the time variable and solve the super-Hamiltonian constraint with respect to it, leading to a reduced Hamiltonian. We now present the two possible extensions of the current formulation of General Relativity involved in this work: $f(R)$ gravity theories and Polymer Quantum Mechanics (PQM). The former studies alternative versions of the usual Einstein's equations by modifying the GR action at the classical level. Such alternative formulations would hopefully describe high-curvature regimes. The latter is implemented to study the implications of the discretized nature of space-time arising in quantum descriptions of GR, like Loop Quantum Gravity, as quantum mechanical effects are expected to play an important role during the Planck era of the Universe's history. %
\subsection{f(R) theories of gravity}
The starting point of Modified Gravity theories is that the current formulation of General Relativity may be part of a more general theory that can also describe high-curvature and high-energy-density regimes. Such more general theories are studied by modifying the usual Einstein-Hilbert action of GR, including higher-order curvature invariants. In particular, a straightforward generalization of GR is to replace the Ricci scalar $R$, present in the Einstein-Hilbert action, with a generic function of it, say $f(R)$. Therefore, the modified action in the vacuum would read as 
\begin{equation}
     S_{f(R)}=-\frac{1}{2k}\int \sqrt{-g}f(R)  \ d^4x \ ,
     \label{eq:HIfR}
\end{equation}
where  $k=8\pi G/c^4$ is the Einstein constant. At this stage, the only requirement on $f(R)$ is that $f(R) \rightarrow R$ in those regimes where GR has already been proven to provide a satisfactory description of experimental observations. Before continuing, it is important to mention that there are at least two variational principles that one can apply to the action in Eq. (\ref{eq:HIfR}): the metric variation and the Palatini variation \cite{Sotiriou_2010}. Even though these two choices yield the same equations of motion when applied to the usual E-H action, this is no longer true in modified gravity. Therefore, from now on, we will consider all the computations to be performed using the metric formalism of f(R) theories. Returning to the action in Eq. (\ref{eq:HIfR}), a convenient way to study these models is to encode the scalar freedom related to the function $f(R)$ into a dynamical scalar field $\chi$. Hence, as long as $f''(\chi) \neq 0$ (here the prime denotes the derivative with respect to $\chi$), the following action 
\begin{equation}
    \begin{aligned}
        S_{f(R)}=-\frac{1}{2k}\int  \sqrt{-g}\left[f(\chi)+f'(\chi)(R-\chi)\right] \ d^4x \ ,
    \end{aligned}
    \label{eq:actionEq}
\end{equation}
is dynamically equivalent to that in Eq. (\ref{eq:HIfR}). Furthermore, we can recast the action above in a Brans-Dicke form via the field redefinition $\xi=f'(\chi)$, with $\xi$ being referred to as the intrinsic scalar field of the theory. Such a redefinition yields the action
\begin{equation}
    \begin{aligned}
        S_{f(R)}=-\frac{1}{2k}\int  \sqrt{-g}\left[\xi R- U(\xi)\right] \ d^4x \ ,
    \end{aligned}
    \label{eq:actionJordan}
\end{equation}
where $U(\xi)=\chi(\xi)\xi-f[\chi(\xi)]$, that is indeed the action of Brans-Dicke theory with Brans-Dicke parameter $\omega_0=0$. This is the so-called {\it Jordan frame} representation of metric $f(R)$ gravity theory. Let us stress the fact that, in this frame, the scalar field, or some function of it, multiplies the Ricci scalar in the lagrangian while matter is typically minimally coupled to the metric. As a result, test particles move on geodesics, but the field equations are very different from Einstein's equations. Regarding the physical nature of the intrinsic scalar field, see \cite{Moretti:2019yhs}. As is usually performed in Brans-Dicke theories, it is possible to obtain another representation of the metric $f(R)$ theory, namely the {\it Einstein frame}, by implementing the conformal transformation $\tilde{g}_{\mu\nu} = \xi\, g_{\mu\nu}$. This transformation leads to the action of a Brans-Dicke theory with $\omega_0=-3/2$. The peculiarity of this frame is that the intrinsic scalar field is now minimally coupled to the metric, but matter does not follow the usual geodesic motion. With the aim of providing a quantum mechanical description of a modified gravity system, we restrict ourselves to the Jordan frame, since the geometrical degrees of freedom and the intrinsic scalar field are independent of each other, allowing a clearer and simpler quantization procedure. 
\subsection{Polymer Quantum Mechanics}
Polymer Quantum Mechanics is a representation of Quantum Mechanics that is unitarily non-equivalent to the standard Schr\"odinger one \cite{Corichi:2007tf}. Its relevance comes from the fact that the effective formulation of Loop Quantum Cosmology, i.e., the application of LQG to cosmological models, is isomorphic to the implementation of PQM to mini-superspace variables \cite{Ashtekar_2015_effective,singh_2005_effective}. The fundamental feature of this theory is that one of the generalized coordinates is regarded as discrete. As a consequence, its conjugate variable is not well defined, but it is possible to define an approximated version of it. The starting point is to consider a set of abstract states $\{\ket{\mu}\},\ \mu \in \mathbb{R}$, for which the inner product between each other yields 
\begin{equation}
    \braket{\mu}{\nu} = \delta_{\mu,\nu}\,,
    \label{eq:abstract-inner}
\end{equation}
with $\delta_{i,j}$ being the Kronecker delta. Let $\hat{q}$ and $\hat{p}$ be the operators related to a configuration space variable and its conjugate momentum, respectively, satisfying the Canonical Commutation Relations $[\hat{q},\hat{p}]=i\hbar \hat{1}$. Relying on the algebraic approach to Quantum Mechanics and the theory of $C^\star$-algebras, we introduce the exponentiated version of these operators in the form
\begin{equation}
    \hat{U}(\alpha)=e^{i \alpha \hat{q}/\hbar}\,, \qquad \hat{V}(\beta)=e^{i \beta \hat{p}/\hbar}\,.
\end{equation}
Here we are only interested in the action of those operators in the momentum polarization, i.e., on those states in the form $\psi_\mu(p) \equiv \braket{p}{\mu} = \text{Exp}[i\mu p/\hbar]$. In this representation, the operator $\hat{V}(\lambda)$ acts multiplicatively as
\begin{equation}
    \hat{V}(\lambda)\, \psi_\mu(p) = e^{i(\mu + \lambda) p /\hbar} = \psi_{\mu + \lambda} (p)\,.
\end{equation}
 Although the label $\mu$ of the elementary states $\psi_\mu(p)$ can take values in a continuum, it has to be regarded as a discrete set in the light of Eq. (\ref{eq:abstract-inner}). In fact, the action of $\hat{V}(\lambda)$ is discontinuous with respect to $\lambda$, since $\ket{\mu}$ and $\ket{\mu+\lambda}$ are always orthogonal regardless of the choice of $\lambda$ \cite{Corichi:2007tf}. Given this feature, the operator $\hat{V}(\lambda)$ is well-defined, but it cannot be generated via the exponentiation of a Hermitian operator and, consequently, $\hat{p}$ is ill-defined. Moreover, as the continuity of the operators $\hat{U}(\alpha),\hat{V}(\lambda)$ is one of the hypotheses of the Stone-von Neumann theorem, the representation we are considering is non-equivalent to the standard Schr\"odinger one. On the other hand, the operator $\hat{q}$ is well-defined and its action reads as
\begin{equation}
    \hat{q}\, \psi_\mu(p) =  i\hbar \partial_p\, \psi_\mu(p) = \mu\, \psi_\mu(p).
\end{equation}
 From the equation above, we can state that $\hat{q}$ is discrete, since its eigenvalues $\mu$ label a discrete set in the sense discussed previously. In light of the representation theory of $C^\star$-algebras, the configuration space on which $p$ lives is the Bohr compactification of the real line $\mathbb{R}_b$ \cite{Corichi:2007tf} and the polymer Hilbert space $\mathcal{H}_{poly}$ is isomorphic to the space
\begin{equation}
    \mathcal{H}_{poly} = L^2(\mathbb{R}_b, d\mu_H)\,,        
\end{equation}
with $d\mu_H$ being the Haar measure. It is possible to recover the notion of the operator $\hat{p}$ by approximating it on a regular graph $\gamma_{\mu_0}$, defined as $\gamma_{\mu_0}=\{q \in \mathbb{R}\ |\ q = n \mu_0,\ \forall n \in \mathbb{Z}\}$. Hence, the momentum operator is well approximated by the incremental ratio
\begin{equation}
    \hat{p} \approx \hat{p}_{\mu_0} = \frac{\hbar}{2i\mu_0} \left[ \hat{V}(\mu_0) - \hat{V}(-\mu_0) \right]\,,
\end{equation}
as long as $p \ll \hbar/\mu_0$. In the momentum space, this operator acts multiplicatively and leads to the formal substitution 
\begin{equation}
    p \to \frac{\hbar}{\mu_0} \sin{(\mu_0 p/\hbar)}\,,
    \label{eq:formalsub}
\end{equation}
with $p \in [-\pi/\mu_0\,,\, \pi/\mu_0]$. In our analysis, the above substitution will be implemented both in the classical equation of motion of the system, to study the polymer semiclassical evolution, and in the WDW equation in momentum representation. Let us also highlight the fact that the inner product for momentum polarization states defined over the graph $\gamma_{\mu_0}$ reduces to the form
\begin{equation}
    \braket{\phi}{\psi}_{\mu_0} = \frac{\mu_0}{2\pi \hbar}\int_{-\pi/\mu_0}^{\pi/\mu_0} dp\ \phi^*(p)\, \psi(p)\,,
    \label{poly_inner}
\end{equation}
that is equivalent to the inner product over the circle $S^1$ \cite{Corichi:2007tf}.
\section{Polymer semiclassical dynamics}\label{sec:semiclassical}
\begin{figure*}
    \centering
    \begin{subfigure}{0.45\textwidth}
        \centering
        \includegraphics[width=1.\textwidth]{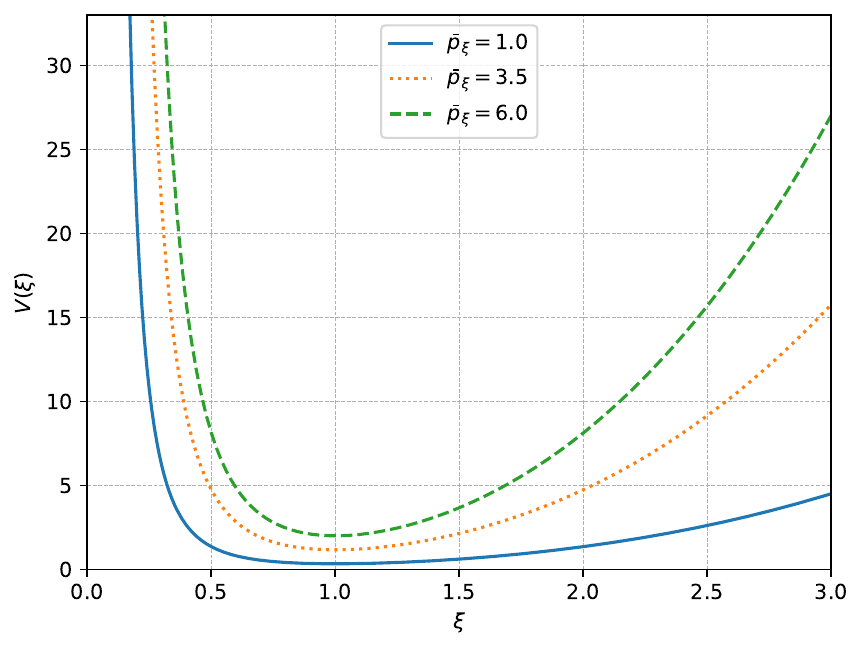}
        \caption{$V(\xi)$}
        \label{fig:1a}
    \end{subfigure}%
    \hfill
    \begin{subfigure}{0.45\textwidth}
        \centering
        \includegraphics[width=1.\textwidth]{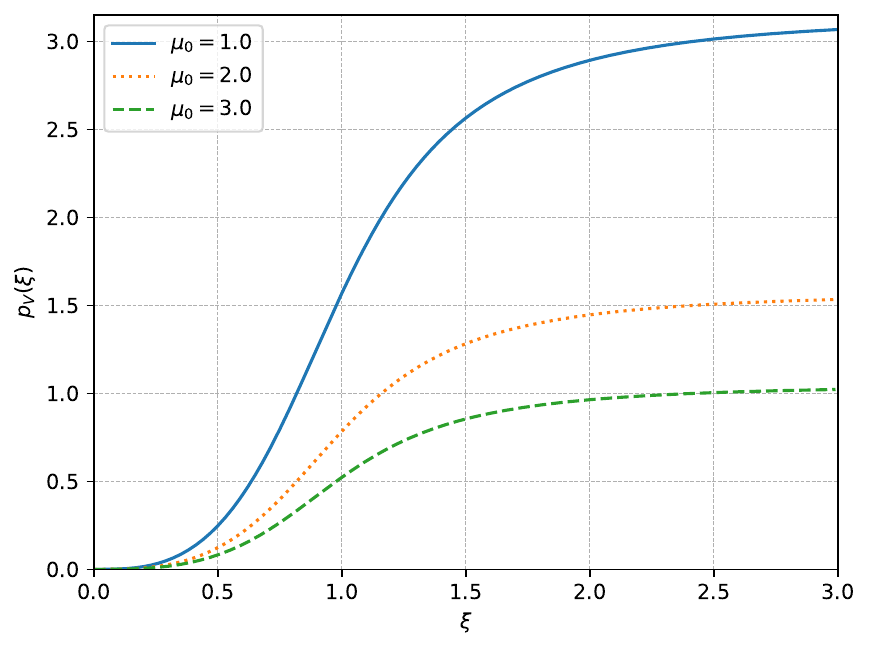}
        \caption{$p_V(\xi)$}
        \label{fig:1b}%
    \end{subfigure}
    \hfill
   % \begin{subfigure}{0.45\textwidth}
   %     \centering
   %     \includegraphics[width=1.\textwidth]{fig1c.pdf}
   %     \caption{$p_\xi(\xi)$}
   %     \label{fig:1c}%
   % \end{subfigure}
    \caption{Plot of semiclassical evolution laws: $V(\xi)$ and $p_V(\xi)$. For $V(\xi)$, we considered different values of the constant $\tilde{p}_\xi$, setting $\mu_0=1$. For $p_V(\xi)$ we considered different values of $\mu_0$.}
    \label{fig:1}%
\end{figure*}
\begin{figure}
    \centering
    \includegraphics[width=.5\columnwidth]{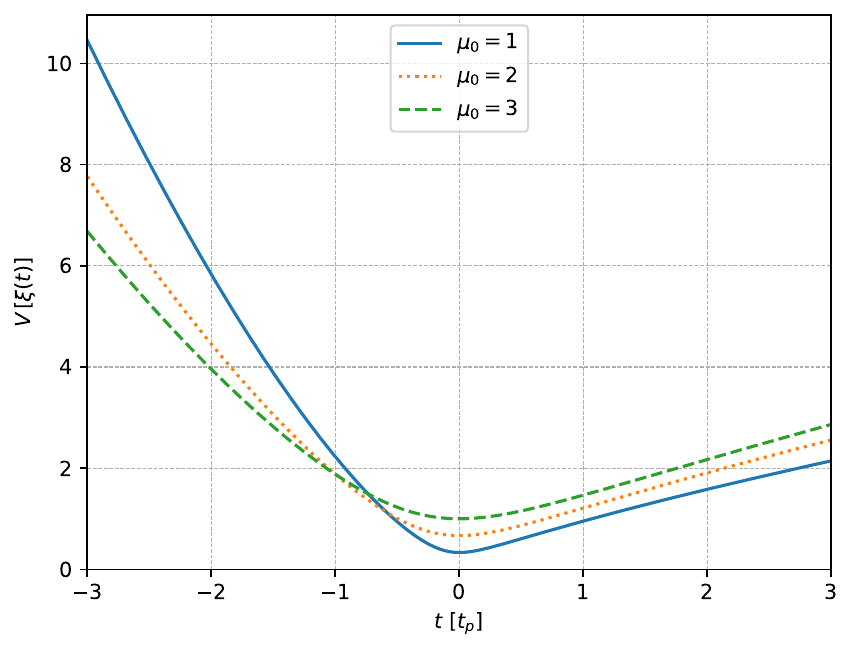}
    \caption{Semiclassical evolution of the volume with respect to the coordinate time in the synchronous reference frame. Different values of $\mu_0$ are considered while fixing $\tilde{p}_\xi=1$.}
    \label{fig:2}%
\end{figure}
\begin{figure}
    \centering
    \includegraphics[width=.5\columnwidth]{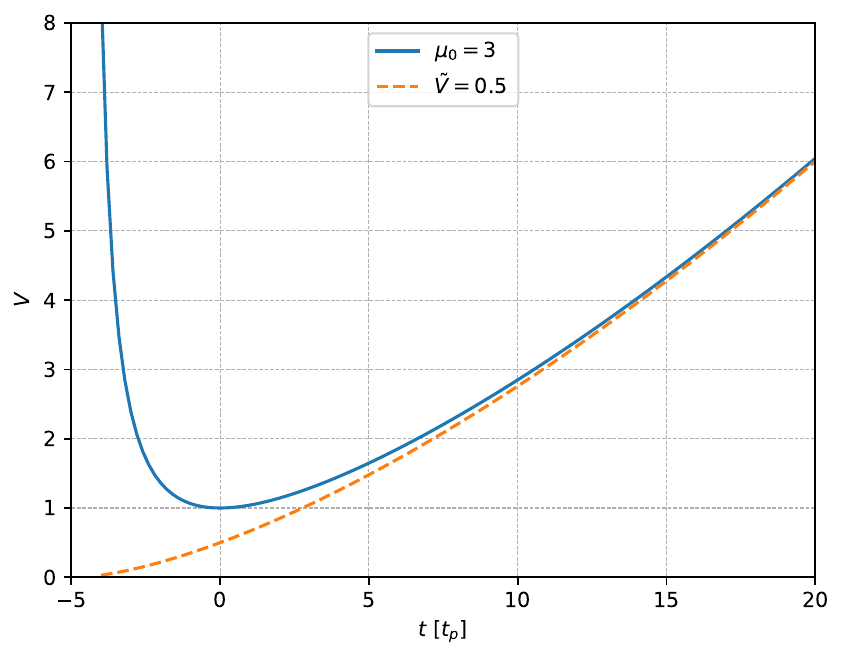}
    \caption{ Semiclassical polymer evolution law $V[\xi_{cl}(t)]$ (solid line) compared with the classical $V_{cl}[\xi_{cl}(t)]$ (dashed line) in the synchronous reference frame. In both cases, we fixed $\tilde{p}_\xi=-1$.}
    \label{fig:3}
\end{figure}
In this section, we compute the semiclassical dynamics of the general FLRW model by substituting the polymer approximation of the momentum operator, i.e., Eq. (\ref{eq:formalsub}), in the modified classical action of Eq. (\ref{eq:HIfR}). In this system, the degree of freedom being discretized will be the volume coordinate $V=a^3(t)$, as suggested by LQC. We start by considering the Jordan frame modified action in eq. (\ref{eq:actionJordan}), and we use the ADM formalism to construct the Hamiltonian of this system. As we are dealing with a spatially homogeneous cosmology, the only non-trivial constraint is the {\it super-Hamiltonian} one, i.e., $\mathcal{H}_{\text{JF}}=0$, where
\begin{equation}
    \mathcal{H}_{\text{JF}}=\frac{k}{6\pi^2}\frac{p_{\xi}^2}{a^3}\xi-\frac{k}{6\pi^2}\frac{p_a p_{\xi}}{a^2}-K\frac{6 \pi^2}{k}\xi a -\frac{\pi^2}{k}a^3 U(\xi)\,.
    \label{eq:superH_1}
\end{equation}
Here $p_a,p_\xi$ are the conjugate momenta of $a, \xi$, respectively. We recall that $\xi>0$ is the intrinsic scalar field and $U(\xi)$ the potential term fixed by the particular choice of the function $f(R)$. Instead, the third term appearing in the above equation is the potential induced by the spatial curvature of the model. Since we are interested in the early Universe dynamics, i.e., close to the initial singularity placed at $a(t) \to 0$, we neglect both these terms in the following derivations. We now describe the systems in terms of the volume coordinate $V$ using the canonical transformation $V=a^3,\,  p_V=p_a/(3a^2)$, and the super-Hamiltonian constraint now reads as
\begin{equation}
    \mathcal{H}_{\text{JF}} = \frac{k}{2\pi^2} \left( \frac{\xi\, p_\xi^2}{3V} - p_\xi p_V\right) = 0\,.
    \label{eq:superH_xi_v}
\end{equation}
In order to study the semiclassical dynamics provided by the discrete character of $V$, we replace $p_V$ with its polymer approximated version using Eq. (\ref{eq:formalsub}). As the LQG formulation of $f(R)$ theories in the Jordan frame has not yet been assessed, given that the choice of Ashtekar variables seems to be affected by the formalism adopted \cite{Bombacigno:2019nua}, here we do not consider the polymerization of the instrinsic scalar field $\xi$. Thus, working in units $\hbar=c=G=1$, we now have
\begin{equation}
    \mathcal{H}_{\text{JF}} = \frac{4}{\pi} \left[ \frac{ \xi\,p_\xi^2}{3V} - \frac{p_\xi}{\mu_0} \sin{(\mu_0 p_V)}\right] = 0\,.
    \label{eq:superH_xi_poly}
\end{equation}
From the Hamiltonian $H_{\text{JF}}=N\mathcal{H}_{\text{JF}}$ we can compute the Hamilton equations of the system, i.e.,
\begin{subequations}
    \begin{equation}
        \dot{V} = -\frac{4N}{\pi} p_\xi \cos{(\mu_0 p_V)}\,, \quad \dot{p}_V = \frac{4N}{3\pi} \frac{\xi\,p_\xi^2}{V^2}\,,
        \label{eq:ham_v_pv_xi}%
    \end{equation}
    \begin{equation}
        \dot{\xi} = \frac{4N}{3\pi} \frac{\xi\, p_\xi}{V}\,, \quad \dot{p}_\xi = -\frac{4N}{3\pi} \frac{p_\xi^2}{V}\,,
        \label{eq:ham_xi_pxi_xi}%
    \end{equation}
    \label{eq:ham_eqs_xi_t}%
\end{subequations}
where the dot denotes the derivative with respect to the coordinate time $t$,  and we used the super-Hamiltonian constraint $\mathcal{H}_{\text{JF}}=0$ to obtain the specific form of the $\dot{\xi}$ equation appearing in Eqs. (\ref{eq:ham_xi_pxi_xi}). From the Eqs. (\ref{eq:ham_xi_pxi_xi}), we can notice that the quantity $\tilde{p}_\xi \equiv \xi\,p_\xi$ is a constant of the motion. To solve the dynamics in a covariant fashion, we combine the equations above to express the generalized coordinates $V,p_V,p_\xi$ as functions of $\xi(t)$. Therefore, by exploiting the constraint in Eq. (\ref{eq:superH_xi_poly}), we find the following set of equations
\begin{subequations}
    \begin{equation}
        \dv{V}{\xi} = -\frac{1}{\xi}\sqrt{9V^2-\mu^2_0\, \xi^2\,p_\xi^2}\,,
        \label{eq:ham_eq_cov_v}%
    \end{equation}
    \begin{equation}
        \dv{p_V}{\xi} = \frac{p_\xi}{V}\,,
        \label{eq:ham_eq_cov_pv}%
    \end{equation}
    \begin{equation}
        \dv{p_\xi}{\xi} = - \frac{p_\xi}{\xi}\,.
        \label{eq:ham_eq_cov_pxi}%
    \end{equation}
    \label{eq:ham_eq_cov}%
\end{subequations}
Notice that Eqs. (\ref{eq:ham_eq_cov}) are independent of the gauge fixing of $N(t)$. From Eq. (\ref{eq:ham_eq_cov_v}) we can immediately state that $V(\xi) \geq V_{min}\equiv\mu_0\tilde{p}_\xi/3$, as the cutoff effects prevent reaching the singularity placed at $V=0$. The solutions of Eqs. (\ref{eq:ham_eq_cov}) reads as
\begin{subequations}
    \begin{equation}
        V(\xi) = \frac{\tilde{p}_\xi\,\mu_0}{6}(\xi^{3}+\xi^{-3})\,,
        \label{eq:v_xi}%
    \end{equation}
    \begin{equation}
        p_V(\xi)=\frac{2}{\mu_0}\arctan{(\xi^3)}\,,
        \label{eq:pv_xi}%
    \end{equation}
    \begin{equation}
        p_\xi(\xi) = \frac{\tilde{p}_\xi}{\xi}\,,
        \label{eq:pxi_xi}%
    \end{equation}
    \label{eq:semicl_xi}%
\end{subequations}
where we set the initial conditions $V(1)=V_{min}\,,\, p_V(1)=\pi/(2\mu_0),$ and $p_\xi(1)\equiv \tilde{p}_\xi$. Figure \ref{fig:1} shows plots of these evolution laws. First of all, the $V(\xi)$ plot in Fig. \ref{fig:1a} exhibits precisely the bounce signature. The Universe undergoes a contraction phase starting from an arbitrarily large volume at $\xi \to 0$. Then, a minimal volume is reached, and an expansion phase starts. The value of the minimal volume depends on both the Polymer step $\mu_0$ and the constant $\tilde{p}_\xi$. An interesting feature of this model is that the bounce is asymmetric. From the $p_V(\xi)$ plot in Fig. \ref{fig:1b}, instead, an interesting fact arises: the values of $p_V$ span only half of its domain during the whole dynamics, being constrained in the interval $[0,\pi/\mu_0]$. This is a consequence of the fact that the two signs of $p_V$ are redundant, and this particular feature will be exploited in the quantum treatment. For later comparison, we now derive the semiclassical evolution of the volume as seen from the synchronous reference system. The gauge fixing corresponding to such a reference frame is $N(t)=1$. In order to find $V[\xi(t)]$, we only need to solve the first equation from Eqs. (\ref{eq:ham_xi_pxi_xi}) using Eqs. (\ref{eq:v_xi},\ref{eq:pxi_xi}). The result reads as
\begin{equation}
    \xi(t) = \frac{1}{{\pi^{1/4}}} \sqrt{\frac{\sqrt[3]{T(t)}}{3^{2/3} \mu_0} + \frac{32 t - \pi \mu_0}{\sqrt[3]{3\,T(t)}}}\,,
    \label{eq:xi_t}
\end{equation}
where $T(t)=9 \pi^{3/2} \mu_0^3 + 2 \sqrt{3 \mu_0^3 (\pi \mu_0 - 8 t) (7 \pi^2 \mu_0^2 + 1024 t^2 + 32 \pi \mu_0 t)}$, and we fixed the initial condition $\xi(0)=1$. Here, we recall that the coordinate time $t$ is expressed in units of Planck time $tp\equiv\sqrt{\hbar G/c^5}$. In Fig. \ref{fig:2} , we show the behavior of $V[\xi(t)]$ obtained by substituting Eq. (\ref{eq:xi_t}) into Eq. (\ref{eq:v_xi}). As we can see, the asymmetric nature of this bounce also manifests itself in the synchronous reference system. Last, we show that the Polymer evolution law recovers the classical one when sufficiently far from the bounce. In fact, by solving the Hamilton's equations derived from the classical constraint in Eq. (\ref{eq:superH_xi_v}), we find the following relations in the synchronous reference frame
\begin{subequations}
    \begin{equation}
        V_{cl}(\xi)= \tilde{V}\, \xi^{-3}\,,
        \label{eq:V_cl}%
    \end{equation}
    \begin{equation}
        \xi_{cl} (t) = \left(1 - \frac{9\pi}{8} \frac{\tilde{V}}{\tilde{p}_\xi}\, t\right)^{-1/2}\,,
        \label{eq:xi_cl}%
    \end{equation}
    \label{eq:V_xi_cl}%
\end{subequations} 
where $\tilde{V}\equiv V(\xi=1)$, $\tilde{p}_\xi$ is the constant of the motion previously defined, and we fixed the initial conditions such that $\xi_{cl}(t=0)=1$. From Eq. (\ref{eq:V_cl}), it is interesting to notice that only the collapsing branch of Eq. (\ref{eq:v_xi}) solves the dynamics, with respect to $\xi$, in the absence of quantum gravitational effects. Therefore, the expansion branch $\xi^{3}$ emerges due to the polymerization of the gravitational degrees of freedom. Nonetheless, in the synchronous reference frame, $V_{cl}[\xi_{cl}(t)]$ can describe both expansion and contraction trajectories depending on the sign of $\tilde{p}_\xi$, i.e., expansion for $\tilde{p}_\xi<0$ and contraction for $\tilde{p}_\xi>0$. In Fig. \ref{fig:3}, we show the comparison between Eq. (\ref{eq:v_xi}) and Eq. (\ref{eq:V_cl}) with respect to $\xi_{cl}(t)$. As we can see, sufficiently far from the bounce, where quantum effects should be negligible, the dynamics of the polymer Universe is in agreement with the classical evolution, given a proper choice of the initial conditions.  

\subsection{Comparison with an external field description}
\begin{figure*}
    \centering
    \begin{subfigure}{0.45\textwidth}
        \includegraphics[width=1.\textwidth]{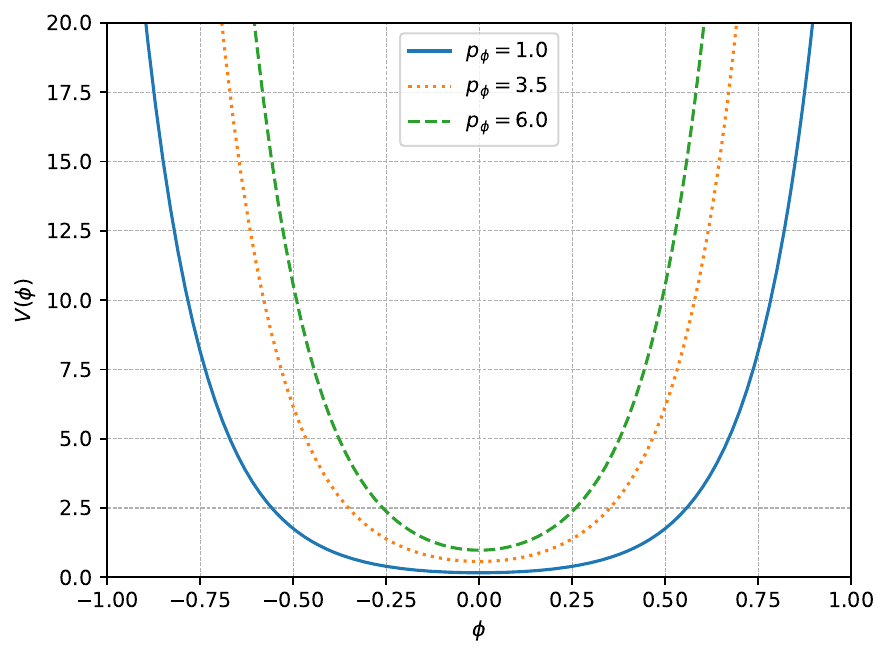}
        \caption{$V(\phi)$}
        \label{fig:4a}%
    \end{subfigure}
    \hfill
    \begin{subfigure}{0.45\textwidth}
        \includegraphics[width=1.\textwidth]{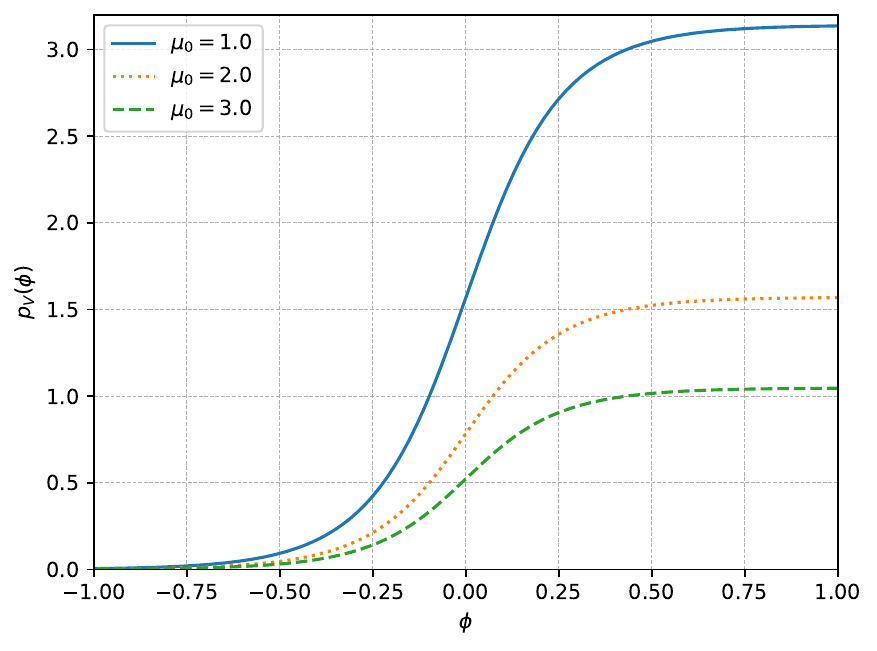}
        \caption{$p_V(\phi)$}
        \label{fig:4b}
    \end{subfigure}
    \caption{Plot of semiclassical evolution laws $V(\phi)$ and $p_V(\phi)$. For $V(\phi)$, we considered different values of the constant $p_\phi$, setting $\mu_0=1$. For $p_V(\phi)$, we considered different values of $\mu_0$.}
    \label{fig:4}%
\end{figure*}
\begin{figure}
    \centering
    \includegraphics[width=.5\columnwidth]{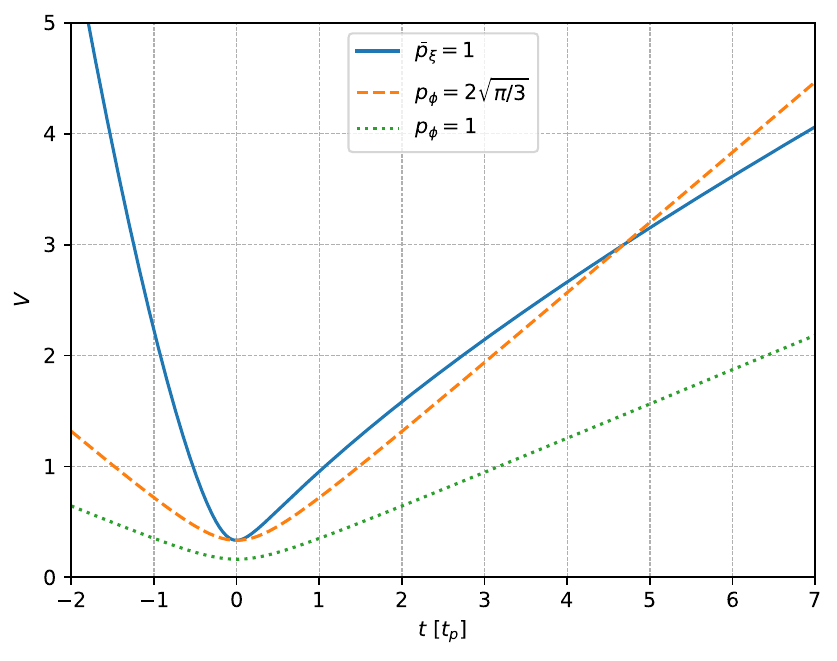}
    \caption{Comparison between semiclassical bounces in the synchronous reference system from $f(R)$ theory (solid line) and standard GR with scalar external field (dashed and dotted lines). Here $\mu_0=1$.}%
    \label{fig:5}
\end{figure}

Now, we want to compare the Bouncing picture derived above with the polymer semiclassical evolution of the same model, considering Einsteinian gravity and the presence of a massless free scalar field. We include such a term as it represents the kinetic term of the inflaton. The main difference from the previous case is that the scalar field is now minimally coupled to the gravitational field. The polymer super-Hamiltonian constraint of this model reads as \cite{Blyth:1975is,Kirillov:2002kc,Montani:PrimCos}
\begin{equation}
    \mathcal{H}_\phi = -\frac{3k}{2}\frac{V}{\mu_0^2} \sin^2{(\mu_0 p_V)} + \frac{p_\phi^2}{V} = 0\,,
    \label{eq:phi_ham_constr}
\end{equation}
where $\phi$ and $p_\phi$ are respectively the scalar field and its conjugate momentum. As stated earlier, we neglected the potential term generated from the spatial curvature. From Eq. (\ref{eq:phi_ham_constr}), we can immediately notice that $p_\phi$ is a constant of the motion. In the same spirit as the previous analysis, we write down the set of covariant equations of motion, reading as
\begin{subequations}
    \begin{equation}
        \dv{V}{\phi} = -\sqrt{12\pi V^2- \mu^2_0\, p_\phi^2}\,,
        \label{eq:ham_eq_cov_v_phi}%
    \end{equation}
    \begin{equation}
        \dv{p_V}{\phi} = \frac{p_\phi}{V}\,,
        \label{eq:ham_eq_cov_pv_phi}%
    \end{equation}
    \begin{equation}
        p_\phi = const\,.
        \label{eq:ham_eq_cov_pphi}%
    \end{equation}
    \label{eq:ham_eq_cov_phi}%
\end{subequations}
where we used the fact that $k=8\pi E_p^2$ in our units. From Eq. (\ref{eq:ham_eq_cov_v_phi}), we can determine the minimal volume mediating this bounce process, i.e., $V_{min}\equiv \mu_0\,p_\phi/(2\sqrt{3\pi})$. In this case too, the minimal volume depends linearly on the polymer step $\mu_0$ and the scalar field conjugate momentum $p_\phi$, but it scales differently from the $f(R)$ case. Therefore, the two models share the same minimal volume when $p_\phi=2\sqrt{\pi/3}\,\tilde{p}_\xi$ for equal polymer steps. The solutions of Eqs. (\ref{eq:ham_eq_cov_phi}) read as
\begin{subequations}
    \begin{equation}
        V(\phi) = \frac{p_\phi\,\mu_0}{2\sqrt{3\pi}}\cosh{\left( 2\sqrt{3\pi}\, \phi\right)}\,,
        \label{eq:v_phi}%
    \end{equation}
    \begin{equation}
        p_V(\phi)=\frac{2}{\mu_0}\arctan{[\tanh{(\sqrt{3\pi}\,\phi)}]}+\frac{\pi}{2\mu_0}\,,
        \label{eq:pv_phi}%
    \end{equation}
    \label{eq:semicl_phi}%
\end{subequations}
fixing the initial conditions $V(0)=V_{min}\,,\,p_V(0)=\pi/(2\mu_0)$. The plots of $V(\phi)$ and $p_V(\phi)$ are shown in Fig. \ref{fig:4}. We can notice that, in the case of Einsteinian gravity, the bounce process is symmetric with respect to the physical clock. Regarding $p_V(\phi)$, its evolution is comparable with the modified gravity case, spanning only the range $[0,\pi/\mu]$. In order to compare the two bounce descriptions, we move to the synchronous reference frame and compute the function $\phi(t)$. The equation to solve is \begin{equation}
    \dot{\phi} = \frac{1}{2\pi^2}\frac{p_\phi}{V}\,,
    \label{eq:ham_phi_t}
\end{equation}
whose solution reads as
\begin{equation}
    \phi(t)=\frac{1}{2\sqrt{3\pi}} \sinh^{-1}\left( \frac{6t}{\pi \mu_0} \right)\,,
    \label{eq:phi_t}
\end{equation}
where we required that $\phi(0)=0$. Substituting this result in Eq. (\ref{eq:v_phi}), we get the evolution law of the volume for the coordinate time in the synchronous reference system $V[\phi(t)]$, i.e.,
\begin{equation}
    V[\phi(t)] = \frac{p_\phi \mu_0}{2\sqrt{3\pi}} \sqrt{1+\frac{36t^2}{\pi \mu_0^2}}\,.
\end{equation}
As we can see, the bounce process is still symmetric in the synchronous frame. In Fig. \ref{fig:5}, we show a graphical comparison between the polymer semiclassical evolution $V[\xi(t)]$ from $f(R)$ theories and $V[\phi(t)]$ from standard GR. As already stated, the main difference between the two descriptions is that the former yields an asymmetric bounce description while the latter a symmetric one. It is interesting to notice that, when the two models share the same minimal volume, the $f(R)$ description is characterized by a steeper contraction. However, during the expansion phase, after starting a more rapid expansion, the $f(R)$ expansion rate drops as the synchronous time grows, and the Einsteinian model overtakes it. To conclude this analysis, we have so far shown that discreteness effects lead to a bouncing cosmology of the FLRW model, also when $f(R)$ modified gravity theories are considered. The resulting scenario is different in several aspects from the standard GR one, both with respect to the physical clock and the synchronous time evolution. In this regard, we remark on the fact that we are investigating the regime in which the higher-order curvature corrections from $f(R)$ theories are relevant; therefore, deviations from standard GR are indeed substantial. In this sense, one would expect the Modified Gravity description to recover the Einsteinian dynamics outside of this region, where also the potential contribution in the complete super-Hamiltonian constraint in Eq. (\ref{eq:superH_1}) cannot be neglected.
\subsection{ADM reduction of the dynamics}
\begin{figure*}
    \centering
    \begin{subfigure}{0.45\textwidth}
        \centering
        \includegraphics[width=1.\textwidth]{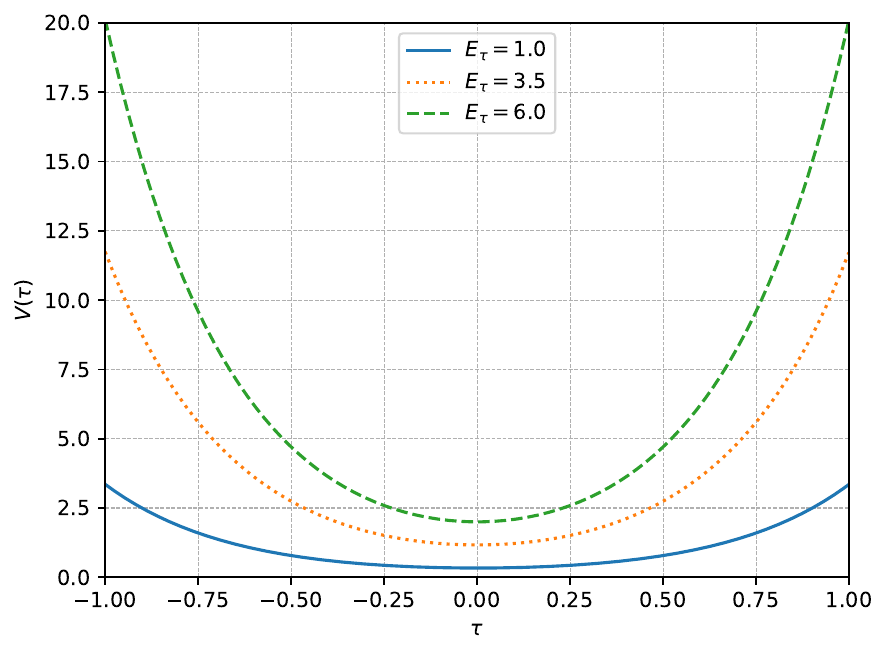}
        \caption{$V(\tau)$}
        \label{fig:6a}%
    \end{subfigure}
    \hfill
    \begin{subfigure}{0.45\textwidth}
        \centering
        \includegraphics[width=1.\textwidth]{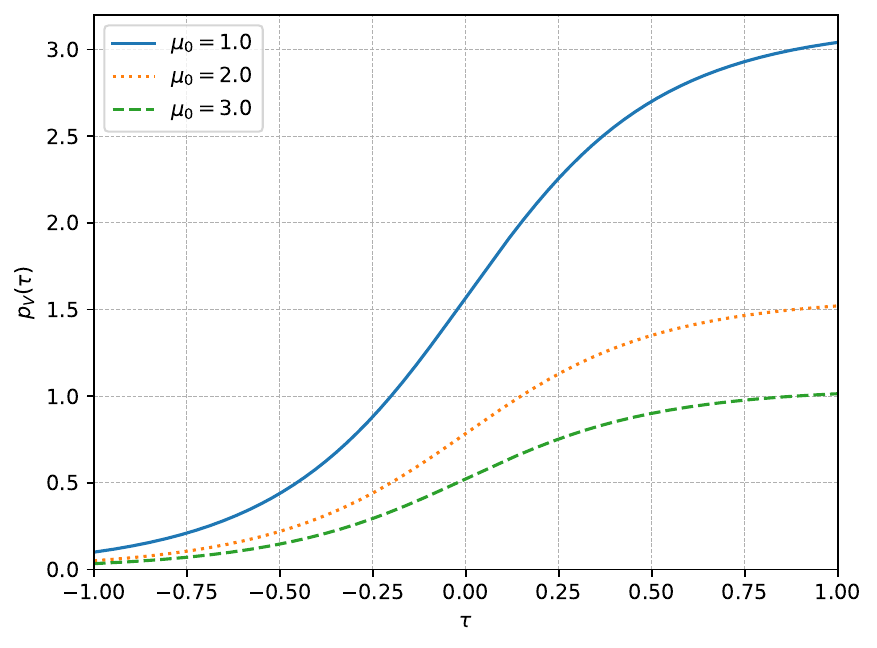}
        \caption{$p_V(\tau)$}
        \label{fig:6b}%
    \end{subfigure}
    \hfill
    \begin{subfigure}{0.45\textwidth}
        \centering
        \includegraphics[width=1.\textwidth]{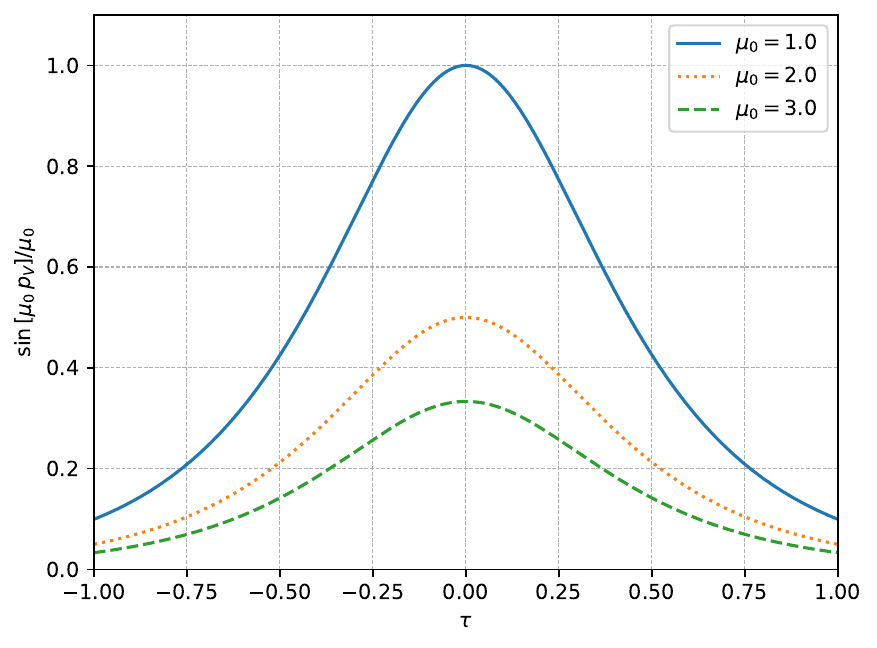}
        \caption{$\sin{[\mu_0p_V(\tau)]}/\mu_0$}
        \label{fig:6c}%
    \end{subfigure}
    \caption{Plot of semiclassical evolution laws $V(\tau)$, $p_V(\tau)$ and $\sin{[\mu_0p_V(\tau)]}/\mu_0$. For $V(\tau)$, we considered different values of the constant $E_\tau$, setting $\mu_0=1$. For $p_V(\tau)$ and $\sin{[\mu_0p_V(\tau)]}/\mu_0$, we considered different values of $\mu_0$.}
    \label{fig:6}
\end{figure*}
Before moving to the quantum mechanical description, here we compute the reduced version of the $f(R)$ dynamical system that will be used in the next section. One of the open issues in the Canonical formulation of Quantum Cosmology is the so-called {\it problem of time}. The vanishing nature of the Hamiltonian of General Relativity yields a Universe's wave function that is independent of the coordinate time $t$, and the notion of time has to be reintroduced at the quantum level. The approach we rely on in this work is to identify one of the system's degrees of freedom to play the role of the time variable, using a {\it time before quantization} procedure. To be more precise, we implement the ADM reduction of the dynamics. Therefore, we solve the super-Hamiltonian constraint (\ref{eq:superH_xi_poly}) at the classical level, and we use the result to obtain the ADM reduced action. For this particular approach, it is convenient to involve the canonical transformation $\tau=\log{\xi}\,,\,p_\tau=\xi\, p_\xi$. Note that the definition of $p_\tau$ coincides with the constant of the motion $\tilde{p}_\xi$ introduced earlier. After this transformation, the super-Hamiltonian constraint in Eq. (\ref{eq:superH_xi_poly}) reads as
\begin{equation}
    \mathcal{H}_{\text{JF}} = \frac{k e^{-\tau}}{2\pi^2} \left[ \frac{p_\tau^2}{3V} - \frac{p_\tau}{\mu_0} \sin{(\mu_0 p_V)}\right] = 0\,.
    \label{eq:superH_tau_poly}
\end{equation}
As we consider $\tau$ the internal time of the system, we solve the eq. (\ref{eq:superH_tau_poly}) in terms of $p_\tau$, obtaining the solutions
\begin{equation}
    p_{\tau,1} = 0\,, \quad p_{\tau,2} = 3V\sin{(\mu_0 p_V)}/\mu_0\,.
\end{equation}
Because the first solution is trivial, we consider only the second one. Fixing the lapse function as $N_{ADM}=3\pi Ve^{\tau}/(4 p_\tau)$, the substitution of $p_{\tau,2}$ into the full expression of the action yields the reduced action
\begin{equation}
    S_{ADM} = \int d\tau\left( p_V \dv{V}{\tau} - h_\tau \right)\,,
    \label{eq:reduced_action}
\end{equation}
where $h_\tau$ is the reduced polymer Hamiltonian
\begin{equation}
    h_\tau = -\frac{3 V}{\mu_0} \sin{(\mu_0 p_V)}\,.
    \label{eq:reduced_ham}
\end{equation}
We can now write the Hamilton equations for $V(\tau)$ and $p_V(\tau)$, i.e.,
\begin{subequations}
    \begin{equation}
        \dv{V}{\tau} = - 3V \cos{(\mu_0 p_V)}\,,
        \label{eq:ham_eqs_v}%
    \end{equation}
    \begin{equation}
        \dv{p_V}{\tau} = 3 \sin{(\mu_0 p_V)}/\mu_0\,.
        \label{eq:ham_eqs_pv}%
    \end{equation}
    \label{eq:ham_eqs}%
\end{subequations}
From Eq. (\ref{eq:ham_eqs_v}), we can notice that, as expected, $V(\tau)$ reaches a minimum when $p_V(\tau) = \pi/(2\mu_0)$ for some value of $\tau$. The solutions of Eqs. (\ref{eq:ham_eqs}) reads as
\begin{equation}
    V(\tau) = V_{min} \cosh{(3\tau)}\,, \quad p_V(\tau) = 2 \arctan{[\exp(3 \tau)]}/\mu_0\,, 
\end{equation}
where we imposed the conditions $V(\tau =0) = V_{min}\,,\, p_V(\tau=0)=\pi/(2\mu_0)$. Since the reduced Hamiltonian (\ref{eq:reduced_ham}) is independent of the internal time explicitly, it represents a constant of the motion, and we can express $V_{min}$ in terms of its value. Therefore, given $h_\tau = -E_\tau$ we find eventually the expressions 
\begin{subequations}
    \begin{equation}
        V(\tau) = \frac{E_\tau\mu_0}{3} \cosh{(3\tau)}\,, 
        \label{eq:v_tau}%
    \end{equation}
    \begin{equation}
        p_V(\tau) = \frac{2}{\mu_0} \arctan{(e^{3 \tau})}\,.
        \label{eq:pv_tau}%
    \end{equation}
    \label{eq:v_pv_tau}%
\end{subequations}
As stated earlier, the energy-like constant $E_\tau$ coincides with that of the covariant treatment, i.e., $\tilde{p}_\xi$. From Eqs. (\ref{eq:v_pv_tau}), we can recover the solutions in Eqs. (\ref{eq:v_xi},\ref{eq:pv_xi}) by substituting $\tau=\log{\xi}$, proving that the ADM reduction provides the same physical description of the dynamics. Moreover, the evolution of $V(\tau)$ has a similar form as that of the Einsteinian case in Eq. (\ref{eq:v_phi}). In Fig. \ref{fig:6}, we show plots of $V(\tau)$ (Fig. \ref{fig:6a}) and $p_V(\tau)$ (Fig. \ref{fig:6b}). As we can see, the dynamics gain a symmetric character with respect to the logarithmic time $\tau$. Here we remark that $\sin{[\mu_0p_V(\tau)]}/\mu_0$ is the relevant physical quantity in the polymer representation, instead of $p_V(\tau)$; Therefore, using Eq. (\ref{eq:pv_tau}), we can obtain the its semiclassical evolution law, reading as
\begin{equation}
    \frac{1}{\mu_0}\sin{[\mu_0p_V(\tau)]} = \frac{1}{\mu_0 \cosh{(3\tau)}} 
    \label{eq:sin_tau}
\end{equation}
In Fig. \ref{fig:6c}, we show the plot of the evolution law from Eq. (\ref{eq:sin_tau}), as it will be used for comparison with results from the quantum framework.  
\section{Polymer quantum dynamics}
We now construct the pure polymer quantum mechanical description of the system under investigation by applying a canonical quantization procedure. Hence, we promote the generalized coordinates $V$ and $p_V$ to operators acting on the Hilbert space of the Universe's wavefunctions. To overcome the problem of time, we directly quantize the reduced Hamiltonian in Eq. (\ref{eq:reduced_ham}), with $\tau$ playing the role of the physical clock. As we work in the polymer momentum representation, the action of the volume operator $\hat{V}$ and of the polymer approximated momentum operator $\hat{p}_{V,\mu_0}$ reads as
\begin{subequations}
    \begin{equation}
        \hat{V}\Psi(p_V)=i\partial_{p_V}\Psi(p_V)\,,
        \label{eq:op_poly_reps_V}%
    \end{equation}
    \begin{equation}
        \hat{p}_{V,\mu_0}\Psi(p_V)= \frac{\sin{(\mu_0p_V)}}{\mu_0} \Psi(p_V)\,.
        \label{eq:op_poly_reps_pv}%
    \end{equation}
    \label{eq:op_poly_reps}%
\end{subequations}
Here, we remark that $V$ is the coordinate exhibiting a discrete character. In the following analyses, we will solve the eigenvalue equation for the reduced Hamiltonian $\hat{h}_\tau$, proving that its solutions form indeed a set of improper eigenstates. These states are used to construct Gaussian wave packets, for which mean values of $\hat{V}$ and $\hat{p}_{V,\mu_0}$ are computed, providing a comparison with semiclassical analysis of the previous section.
\subsection{The Hamiltonian operator and its eigenstates}
By promoting generalized coordinates to operators, the reduced Hamiltonian in Eq. (\ref{eq:reduced_ham}) becomes an operator in the form
\begin{equation}
    \hat{h}_\tau = -\frac{3}{2} \left[ \hat{V} \hat{p}_{V,\mu_0} + \hat{p}_{V,\mu_0}\hat{V}\right]\,,
    \label{eq:ham_oper_hat}
\end{equation}
where we choose a symmetric operatorial ordering. It is interesting to notice that the Hamiltonian operator written in this form corresponds to the polymer version of the dilatation operator up to a constant. In terms of the representations in Eq. (\ref{eq:op_poly_reps}), the Hamiltonian operator reads as
\begin{equation}
    \begin{split}
    \hat{h}_\tau = - i\frac{3}{2\mu_0}\left[ \partial_{p_V}\sin{(\mu_0p_V)} + \sin{(\mu_0p_V)}\partial_{p_V} \right]\,.
    \end{split}
    \label{eq:ham_oper_diff}
\end{equation}
As stated in \cite{Corichi:2007tf}, when working in the polymer regular graph, the Hilbert space on which these operators act is the space of square integrable functions on $S^1$, equipped with the inner product in Eq. (\ref{poly_inner}). Moreover, we restrict our analysis to the half-circle, so that our Hilbert space is $\mathscr{H} \coloneq L^2([0,\pi/\mu_0],dp_V)$. This choice is justified by the fact that the two half-circles represent two independent branches of the system, as can be seen immediately from the semiclassical dynamics where $p_V$ is constrained on the interval $[0,\pi/\mu_0]$. Later, we will provide another argument in favor of this restriction. Using tools from the analysis of unbounded operators \cite{Hall2013}, it is possible to show that $\hat{h}_\tau$ is a symmetric operator (hermitian and dense) and its deficiency indices are both zero; therefore, it admits a unique self-adjoint extension that is the adjoint $\hat{h}^\dagger_\tau$, i.e.,
\begin{equation}
    \hat{h}^\dagger_\tau = - i\frac{3}{2\mu_0}\left[ D_{p_V}\sin{(\mu_0p_V)} + \sin{(\mu_0p_V)}D_{p_V} \right]\,.
    \label{eq:ham_oper_adjo}
\end{equation}
Here $D_{p_V}$ denotes the weak derivative. The domain of this self-adjoint extension is 
\begin{equation}
    \mathcal{D}(\hat{h}^\dagger_\tau) \coloneq \left\{ \psi(p_V) \in \mathscr{H} \mid \hat{h}^\dagger_\tau\, \psi(p_V) \in \mathscr{H} \right\}\,,
    \label{eq:ham_oper_adjo_dom}
\end{equation}
where no precise boundary conditions are involved. Before moving on, it is important to discuss the self-adjointness of the differential representation of the volume operator in Eq. (\ref{eq:op_poly_reps}), as it will be used later to compute its mean value. Because our wavefunctions have support on a closed interval, the operator $\hat{V}$ admits an infinite number of self-adjoint extensions, whose domains are
\begin{equation}
    \mathcal{D}(\hat{V}_\theta) \coloneq \left\{ \psi(p_V) \in \mathscr{H} \mid \psi(0) = e^{i\theta} \psi(\pi/\mu_0) \right\}\,.
    \label{eq:v_slfadj_dom}
\end{equation}
Since the domain of the Hamiltonian operator does not impose additional conditions, we decide to deal with the set of functions characterized by strictly periodic boundary conditions, i.e., $\psi(0)=\psi(\pi/\mu_0)$, corresponding to the $\theta=0$ self-adjoint extension. For this choice, the action of the week derivative $D_{p_V}$ coincides with that of the ordinary derivative $\partial_{p_V}$. Therefore, from now on, when dealing with both the Hamiltonian and volume operators, we implicitly refer to their self-adjoint extensions. Now, let us consider the eigenvalue equation of the Hamiltonian operator $\hat{h}_\tau\, \psi_\omega = -\omega\, \psi_\omega$, that reads as
\begin{equation}
    i\frac{3}{2\mu_0}\left[ \partial_{p_V}\sin{(\mu_0p_V)} + \sin{(\mu_0p_V)}\partial_{p_V} \right]\, \psi_\omega =\omega\, \psi_\omega\,.
    \label{eq:eigenv_ham}
\end{equation}
Note that the minus sign in front of the Hamiltonian eigenvalue is a convention to make contact with the energy-like constant $E_\tau$ introduced in the previous section. The solutions of this equation are
\begin{equation}
    \psi_\omega (p_V) = C\frac{\tan{(\mu_0 p_V/2)}^{-i \omega /3}}{\sqrt{\sin{(\mu_0 p_V)}}}\,,
    \label{eq:ham_eigenstates}
\end{equation}
where $C$ is a normalization constant. We can immediately notice that these functions are not square integrable on the interval $[0,\pi/\mu_0]$. Hence, we are going to show that the solutions in Eq. (\ref{eq:ham_eigenstates}) form a set of improper eigenstates by proving the orthonormality relation $\braket{\psi_{\omega'}}{\psi_\omega} = \delta(\omega-\omega')$, with $\delta(\cdot)$ being the usual Dirac delta. The inner product between two different solutions reads as
\begin{equation}
    \braket{\psi_{\omega'}}{\psi_\omega} = \frac{C^2\mu_0}{2\pi}\int_0^{\frac{\pi}{\mu_0}} dp_V\, \frac{\tan{(\mu_0 p_V/2)}^{-\frac{i}{3} (\omega-\omega')}}{\sin{(\mu_0 p_V)}}.
    \label{eq:eigens_inner_1}
\end{equation}
By implementing the variable transformation $u=\tan{(\mu_0p_V/2)}$, the integral above can be recasted in the form
\begin{equation}
    \braket{\psi_{\omega'}}{\psi_\omega} = \frac{C^2}{2\pi}\int_0^{+\infty} du\ u^{-1-\frac{i}{3}(\omega-\omega')}\,.
    \label{eq:eigens_inner_2}
\end{equation}
With the help of the additional transformation $s=\log u$, eventually, we obtain the integral
\begin{equation}
    \braket{\psi_{\omega'}}{\psi_\omega} = \frac{C^2}{2\pi}\int_{\mathbb{R}} ds\ e^{-i\frac{\omega-\omega'}{3} s} = 3C^2\delta(\omega-\omega')\,.
    \label{eq:eigens_inner_3}
\end{equation}
By choosing $C=1/\sqrt{3}$, the desired orthonormality relation holds. Therefore, the normalized solutions read as
\begin{equation}
    \psi_\omega (p_V) = \frac{\tan{(\mu_0 p_V/2)}^{-i \omega /3}}{\sqrt{3\sin{(\mu_0 p_V)}}}\,,
    \label{eq:ham_eigenstates_norm}
\end{equation}
The fact that the functions in Eq. (\ref{eq:ham_eigenstates_norm}) form a set of improper eigenvalues confirms that the Hamiltonian is indeed a self-adjoint operator and its eigenstates can be used to construct localized wavepackets that are in $L^2([0,\pi/\mu_0],dp_V)$. The need for restricting to the half-circle $[0,\pi/\mu_0]$ can also be justified by evaluating the inner product $\braket{\psi_{\omega'}}{\psi_\omega}$ on the other half, i.e., $[-\pi/\mu_0,0]$. In fact, by involving similar transformations to those shown above, we get that 
\begin{equation}
    \frac{\mu_0}{2\pi}\int^0_{-\frac{\pi}{\mu_0}} dp_V\ \psi^*_{\omega'}(p_V)\, \psi_{\omega}(p_V) = -\delta(\omega-\omega')\,.
    \label{eq:eigens_inner_other_half}
\end{equation}
As stated above, it is clear that the two halves of the circle represent two redundant branches up to a minus sign; as ill-defined probability densities arise in the interval $[-\pi/\mu_0,0]$ due to this minus sign, invoking a redefinition of the inner product, we directly selected the other half to construct our Hilbert space.
\subsection{Gaussian wavepackets}
\begin{figure*}
    \centering
    \begin{subfigure}{0.48\textwidth}
        \includegraphics[width=1.\textwidth]{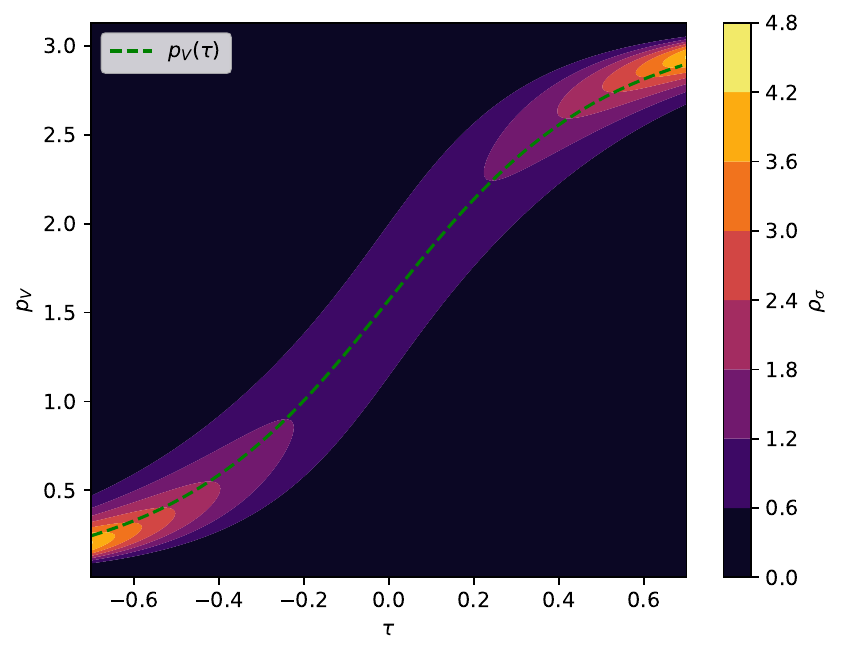}
        \caption{$\sigma=5$}
        \label{fig:7a}
    \end{subfigure}
    \hfill
    \begin{subfigure}{0.48\textwidth}
        \includegraphics[width=1.\textwidth]{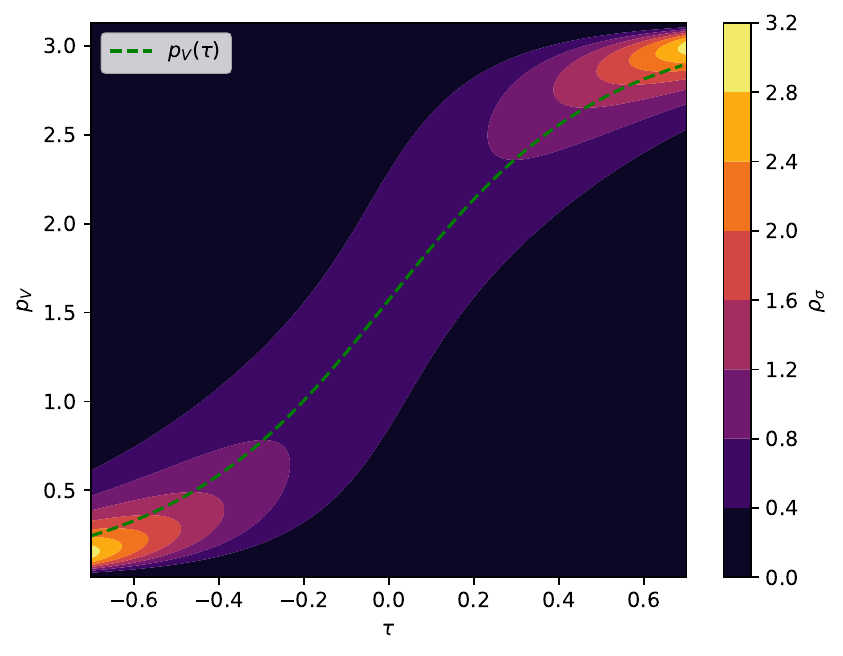}
        \caption{$\sigma=3$}
        \label{fig:7b}
    \end{subfigure}
    \caption{Plots of the probability density $\rho_\sigma(p_V,\tau)$ of Gaussian wave packets. We consider different values of $\sigma$, while fixing $\mu_0=1$, The green dashed line represents the semiclassical evolution law $p_V(\tau)$.}
    \label{fig:7}
\end{figure*}
The internal time evolution of this system is controlled by the Schr\"odinger-like equation 
\begin{equation}
    i\partial_\tau \Psi(p_V,\tau) = \hat{h}_\tau \Psi(p_V,\tau)\,.
    \label{eq:schr-like}
\end{equation}
Given the Hamiltonian improper eigenstates in Eq. (\ref{eq:ham_eigenstates_norm}), the basis solutions of Eq. (\ref{eq:schr-like}) are in the form
\begin{equation}
    \psi_\omega(p_V,\tau) = e^{i\omega\tau} \psi_\omega(p_V)\,,
    \label{eq:schr-sol}
\end{equation}
and they can be used to construct Gaussian localized wave packets as
\begin{subequations}
    \begin{equation}
        \Psi_{\Omega,\sigma}(p_V,\tau) = \int_{\mathbb{R}} d\omega\ A(\omega; \Omega,\sigma)\psi_\omega(p_V,\tau)\,,
        \label{eq:wavepackets_int}
    \end{equation}    
    \begin{equation}
        A(\omega; \Omega,\sigma) = \frac{\exp{\frac{(\omega-\Omega)^2}{2\sigma^2}}}{\sqrt{\sigma\sqrt{\pi}}}\,.
        \label{eq:gaussian_weight}
    \end{equation}
    \label{eq:wavepackets}%
\end{subequations}
Here $\Omega$ and $\sigma$ are respectively the mean value and variance of the Gaussian weight $A(\omega; \Omega,\sigma)$, i.e., the mean value and variance of the Hamiltonian operator. The normalization factors in Eq. (\ref{eq:gaussian_weight}) are chosen to make a unit norm wave packet, i.e., $\braket{\Psi_{\Omega,\sigma}}{\Psi_{\Omega,\sigma}}=1$. The integral in Eq. (\ref{eq:wavepackets_int}) can be solved by analytical means, so that the wavepackets $\Psi_{\Omega,\sigma}$ and their probability densities $\rho\equiv \mu_0 \abs{\Psi_{\Omega,\sigma}}^2/(2\pi)$ are explicitly in the form 
    \begin{subequations}
        \begin{equation}
            \Psi_{\Omega,\sigma}(p_V,\tau) = \sqrt{\frac{2 \sigma\sqrt{\pi}}{3\sin{(\mu_0p_V)}}}\exp{i\Omega\tau-\frac{1}{2}\sigma^2\left[\tau^2+\frac{1}{9}\log{\left[\tan{\left(\frac{\mu_0p_V}{2}\right)}\right]}^2\right]}\tan{\left(\frac{\mu_0p_V}{2}\right)^{\frac{1}{3}(\sigma^2\tau-i\Omega)}}\,,
            \label{eq:wp_exp}%
        \end{equation}
        \begin{equation}
            \rho_{\sigma}(p_V,\tau) = \frac{\mu_0 \sigma}{3\sqrt{\pi}\sin{(\mu_0p_V)}} \exp{-\frac{1}{9}\sigma^2\left[3\tau-\log{\left[\tan{\left(\frac{\mu_0p_V}{2}\right)}\right]}\right]^2}\,.
            \label{eq:rho}%
        \end{equation}
    \end{subequations}
    \label{eq:wp_rho}%

Let us comment a bit on the properties of these quantities. First, we can notice that $\Psi_{\Omega,\sigma}(0,\tau)=\Psi_{\Omega,\sigma}(\pi/\mu_0,\tau)=0$. Hence, these wave packets live in the domain of the $\theta=0$ self-adjoint extension of the volume operator as requested. Trivially, the relation $\rho_{\sigma}(0,\tau)=\rho_{\sigma}(\pi/\mu_0,\tau)=0$ holds. Another interesting fact is that the probability density in Eq. (\ref{eq:rho}) is independent of $\Omega$. As $\Omega$ is the quantum equivalent of the energy-like constant $E_\tau$, we recover the result of the semiclassical analysis in Eq. (\ref{eq:pv_tau}), where the behavior of $p_V(\tau)$ is independent of $E_\tau$. Instead, to clarify the role of $\sigma$, in Fig. \ref{fig:7} we show the evolution of the probability density $\rho_\sigma$ with respect to the internal time $\tau$ for two different values of $\sigma$. As we can see, the probability density is centered around the semiclassical trajectory and it spreads near the instant of the bounce at $\tau=0$, while being narrowly peaked at early/late times. By looking at the differences between Figs. \ref{fig:7a} an \ref{fig:7b}, we can state that $\sigma$ is responsible for the localization of the state. Higher values of $\sigma$, corresponding to less localized states with respect to Hamiltonian eigenvalues, yield a narrower probability density. In particular, the difference in localization with respect to $p_V$ is evident near the turning point of the bounce. Therefore, given this spreading behavior, the validity of the semiclassical bounce picture might depend on the choice of the initial condition $\sigma$.   
\subsection{Mean values and link to the semiclassical analysis}
\begin{figure*}
    \centering
    \begin{subfigure}{0.48\textwidth}
        \includegraphics[width=1.\textwidth]{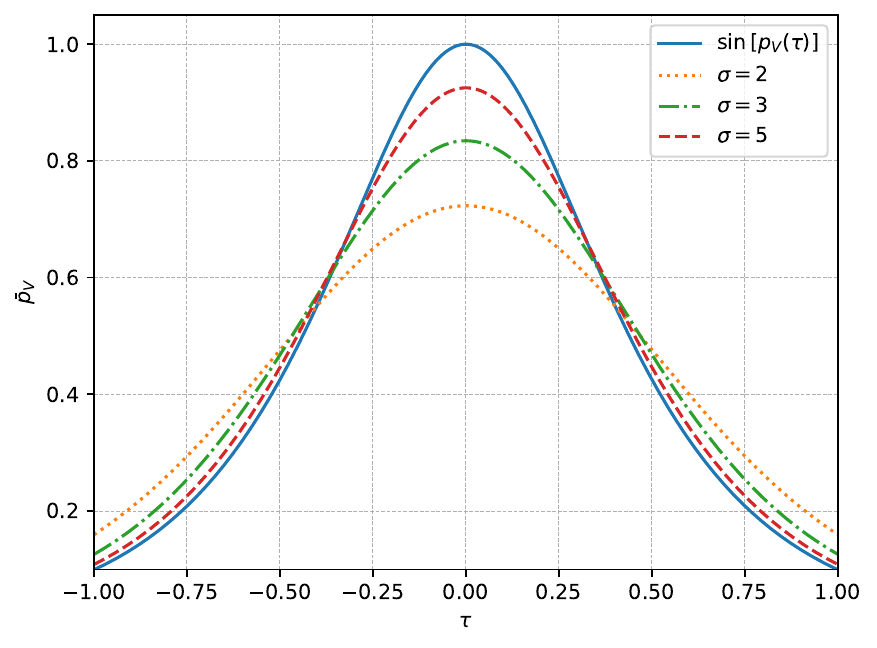}
        \caption{$\bar{p}_V(\tau;\sigma)$}
        \label{fig:8a}
    \end{subfigure}
    \hfill
    \begin{subfigure}{0.48\textwidth}
        \includegraphics[width=1.\textwidth]{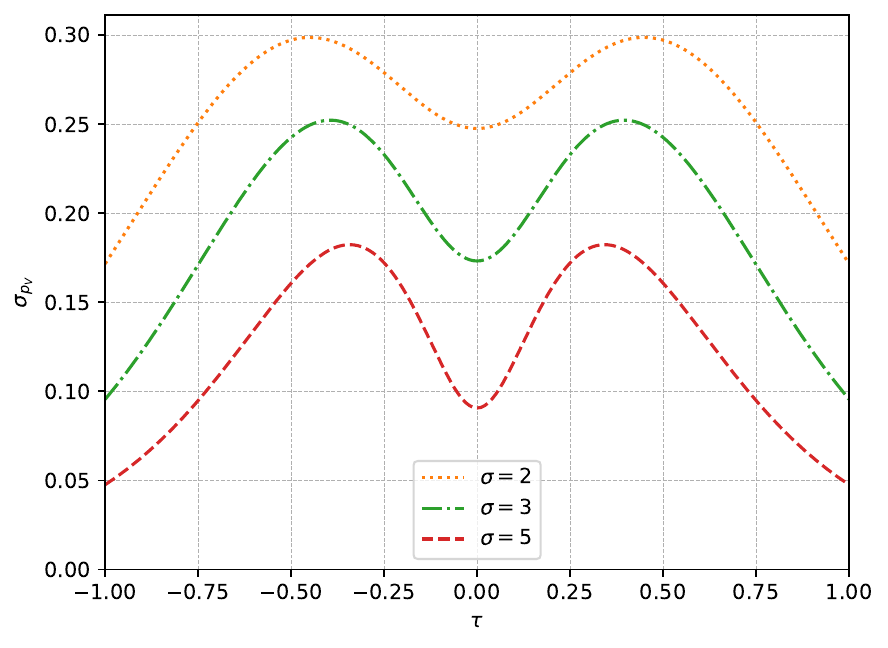}
        \caption{$\sigma_{p_V}(\tau;\sigma)$}
        \label{fig:8b}
    \end{subfigure}
    \caption{Plots of the internal time evolution of the operator $\hat{p}_{V,\mu_0}$ mean value and variance, computed over Gaussian wave packets. Different values of $\sigma$ are considered and $\mu_0=1$. The solid line represents the semiclassical evolution law in Eq. (\ref{eq:sin_tau}).} 
    \label{fig:8}
\end{figure*}
\begin{figure*}
    \centering
    \begin{subfigure}{0.48\textwidth}
        \includegraphics[width=1.\textwidth]{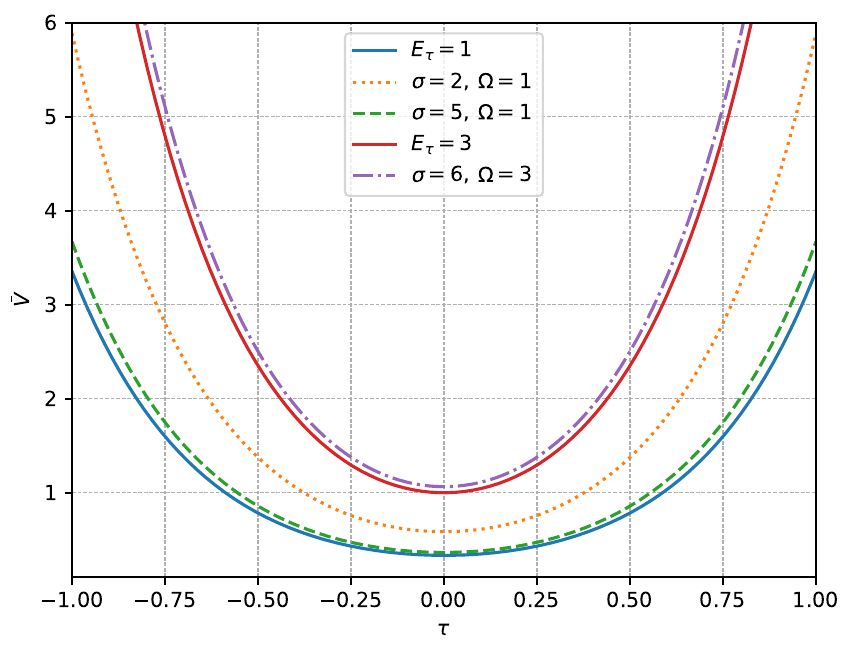}
        \caption{$\bar{V}(\tau;\Omega,\sigma)$}
        \label{fig:9a}
    \end{subfigure}
    \hfill
    \begin{subfigure}{0.48\textwidth}
        \includegraphics[width=1.\textwidth]{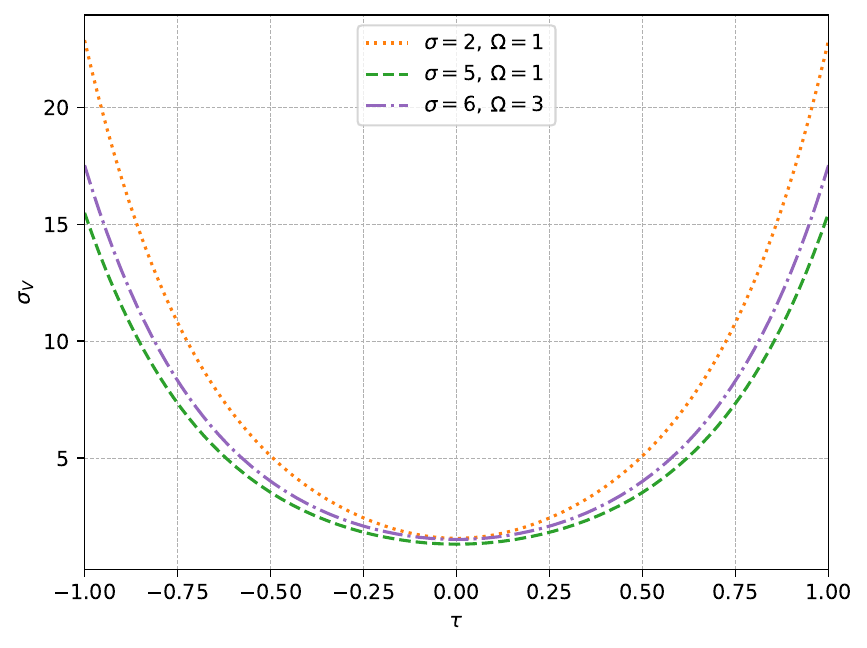}
        \caption{$\sigma_V(\tau;\Omega,\sigma)$}
        \label{fig:9b}
    \end{subfigure}
    \caption{Plots of the internal time evolution of the operator $\hat{V}$ mean value and variance, computed over Gaussian wave packets. Different values of $\sigma$ and $\Omega$ are considered, while $\mu_0=1$. The solid lines represent the semiclassical evolution law in Eq. (\ref{eq:v_tau}) with two different values of $E_\tau$.}
    \label{fig:9}
\end{figure*}

We now compute the internal time evolution of mean values and variances from $\hat{V}$ and $\hat{p}_{V,\mu_0}$. Let us start from the polymer approximated momentum operator; from Eq. (\ref{eq:op_poly_reps_pv}), we know that $\hat{p}_{V,\mu_0}$ acts multiplicatively in the representation we use. Therefore, denoting with $\bar{p}_V\,,\,\sigma_{p_V}$ respectively the mean value and variance of $\hat{p}_{V,\mu_0}$, we find that 
\begin{subequations}
    \begin{equation}
        \bar{p}_V(\tau;\sigma) = \int_{0}^{\frac{\pi}{\mu_0}} \frac{dp_V}{2\pi}\ \sin{(\mu_0p_V)} \rho_\sigma\,,
        \label{eq:pv_mean}%
    \end{equation}
    \begin{equation}
        \sigma^2_{p_V}(\tau;\sigma) = \int_{0}^{\frac{\pi}{\mu_0}} \frac{dp_V}{2\pi\mu_0}\ \sin^2{(\mu_0p_V)} \rho_\sigma- \bar{p}_V^{\,2}\,.
        \label{eq:pv_var}%
    \end{equation}
    \label{eq:pv_mean_var}%
\end{subequations}
Note that both $\bar{p}_V\,,\,\sigma^2_{p_V}$ are independent of $\Omega$ as a straightforward consequence of the specific form of the probability density $\rho_\sigma$. The integrals in Eqs. (\ref{eq:pv_mean_var}) are computed by means of numerical integrations, and in Fig. \ref{fig:8}, we show the internal time evolution of these quantities. Starting from $\bar{p}_V(\tau)$ (Fig. \ref{fig:8a}), we notice that the mean value of $\hat{p}_{V,\mu_0}$ follows the semiclassical evolution in Eq. (\ref{eq:pv_tau}) as long as the value of $\sigma$ is high enough. On the other hand, for smaller values of it, $\bar{p}_V(\tau)$ reproduces with less accuracy the semiclassical dynamics. Regarding the variance $\sigma^2_{p_V}$ (Fig. \ref{fig:8b}), we can see that its value goes to zero at asymptotically early/late times, while taking non-negligible values near the bounce. The parameter $\sigma$ affects its behavior, as expected from what we have learned so far, by yielding an overall greater degree of uncertainty for smaller values of it. Moreover, a non-trivial dependence arises in the $\sigma^2_{p_V}$ evolution, as the variance reaches a local minimum in correspondence with the turning point. Such a minimum is steeper the higher $\sigma$ is chosen. \\ \\ Now, let us compute the mean value and variance of the volume operator. In the momentum representation, $\hat{V}$ acts as a differential operator that has infinite self-adjoint extensions. As previously discussed, we select the $\theta=0$ extension, requiring a strict periodicity condition on the boundaries of the momentum domain, and Gaussian wave packets are suitable to be treated with this particular choice. Therefore, the representation of the volume operator in Eq. (\ref{eq:op_poly_reps_V}) acts as a self-adjoint operator on our states, and its mean value and variance read as
\begin{subequations}
    \begin{equation}
        \bar{V}(\tau;\Omega,\sigma) = i\frac{\mu_0}{2\pi} \int_0^{\frac{\pi}{\mu_0}} dp_V\ \Psi_{\Omega,\sigma}\, \partial_{p_V}\Psi_{\Omega,\sigma}\,,
        \label{eq:v_mean}%
    \end{equation}
    \begin{equation}
        \sigma^2_V(\tau;\Omega,\sigma) = -\frac{\mu_0}{2\pi} \int_0^{\frac{\pi}{\mu_0}} dp_V\ \Psi_{\Omega,\sigma}\, \partial_{p_V}^2\Psi_{\Omega,\sigma}-\bar{V}^2\,.
        \label{eq:v_var}%
    \end{equation}
    \label{eq:v_mean_var}%
\end{subequations}
Notice that now these quantities depend on the Hamiltonian mean value $\Omega$. Also in this case, we use numerical integration techniques to compute the integrals in Eqs. (\ref{eq:v_mean_var}). The results of these computations are shown in Fig. \ref{fig:9}. In Fig. (\ref{fig:9a}), we show the internal time evolution of $\bar{V}$ for different values of $\Omega$ and $\sigma$ and compare it with the semiclassical law in Eq. (\ref{eq:v_tau}). As we can see, we recover the bounce picture at the level of the volume mean value. Regarding the role of the initial conditions, $\Omega$ is mainly responsible for the position of minimal volume, in analogy with the constant $E_\tau$. However, $\sigma$ also affects the evolution, shifting the minimal volume. Therefore, also for the volume operator, we find the same behavior observed for $\bar{p}_V$, as $\bar{V}(\tau;\Omega,\sigma)$ departs from the semiclassical law when choosing small values of $\sigma$. Moving to the $\sigma_V(\tau;\Omega,\sigma)$ plot in Fig. \ref{fig:9b}, we observe that it has a global minimum in correspondence with the bounce, while rapidly increasing at early/late times. Therefore, by comparison with the internal time evolution of the momentum variance in Fig. \ref{fig:8b}, the minimum of overall uncertainty of the system is placed at the bounce instant. Regarding the dependence on initial conditions, it is evident that the volume operator variance is sensitive to the variation of both $\Omega$ and $\sigma$. However, these differences become important only far from the turning point. \\ \\
In light of these considerations, let us summarize the results obtained here and provide an overall commentary. By constructing Gaussian wave packets, using the basis solutions of Eq. (\ref{eq:schr-like}), we were able to show the existence of a bouncing dynamics at the level of $\hat{V}$ and $\hat{P}_{V,\mu_0}$ mean values in the full quantum regime. Therefore, we confirm that the interplay between high-energy classical modifications of GR and cutoff physics effects allows for the removal of the singularity \emph{\`a la} Ashtekar \cite{Ashtekar_2011}, after identifying the additional scalar degrees of freedom from $f(R)$ theories as the physical clock of the system. \\ \\ 
However, a discussion regarding the role of the initial conditions $\Omega$ and $\sigma$ must be carried on, in particular on the latter. Here we recall that they represent the reduced Hailtonian mean value and variance, respectively, based on the definition of our wave packets in Eq. (\ref{eq:wavepackets}). While $\Omega$ behaves like the constant $E_\tau$, as it does not affect the information about $\hat{P}_{V,\mu_0}$ and it controls the minimal volume position for $\bar{V}(\tau;\Omega,\sigma)$, the effects of the variation of $\sigma$ are less trivial, in particular, regarding the holding of the semiclassical approximation. When we consider states that are particularly localized with respect to the Hamiltonian operator, the semiclassical evolution laws in Eqs. (\ref{eq:v_tau},\ref{eq:sin_tau}), fail to represent the behavior of the operators' mean values. This fact, combined with the spreading nature of the probability density typical of the polymer framework and appreciable for such values of $\sigma$, suggests that the validity of the semiclassical bouncing cosmology is dependent on the choice of initial conditions. We postpone to a later analysis the investigation of the deviations from the semiclassical approximation.   
\section{Concluding remarks}
We analyzed the quantum dynamics of an isotropic Universe in the framework of a metric $f(R)$-gravity in the Jordan frame, implementing a polymer quantum mechanics picture and adopting the non-minimal scalar field arising from the theory as a physical clock. This analysis generalized previous studies that polymerized the Universe volume in a standard Einsteinian scheme, as in \cite{Corichi:2007tf,Barca:2021qdn}, or canonically quantized metric $f(R)$-gravity \cite{Bamonti:2021jmg, DeAngelis:2021afq}. \\ \\%
First of all, we analyzed the so-called semi-classical polymer formulation of the model dynamics, i.e., we studied the classical system evolution where the classical Universe is replaced by the polymer-like representation of momentum, in the spirit of the Ehrenfest theorem. To solve the dynamics, we worked within a covariant Hamiltonian formulation. This analysis allowed us to show the semi-classical existence of a bounce, and a comparison with the standard bounce picture was also possible, by moving to the synchronous reference system in both cases. We have seen how the bouncing cosmology has similar features in these two cases; however, when the present modified gravity scenario is addressed, the dynamics is not symmetric with respect to the minimum volume configurations (like in the standard Big-Bounce model \cite{Ashtekar:2005qt}), i.e, the collapsing and expanding branches are not equivalent. \\ \\% 
Then, we studied the full quantum case in the momentum representation, the only one allowed by the polymerization of the model. We constructed a suitable Hilbert space for the dynamics and, hence, we described the evolution of localized wave packets in momentum space. As a result, we were able to describe the mean value behavior for the volume conjugate momentum, which is shown to follow a classical trajectory. Furthermore, choosing the proper self-adjoint extension of the volume operator for the states under consideration, we were able to describe the quasi-classical trajectory of the volume in terms of the non-minimally coupled scalar field, corresponding to a bonce profile, very similar to the original one proposed by the Ashtekar School \cite{Ashtekar_2011}. \\ \\%
The analysis on the choice of initial conditions distinguished between the cases that can be described using a semiclassical formulation. Our model \emph{de facto} corresponds to the implementation of both quantum and Lagrangian corrections to a standard Big-Bang scenario for the isotropic Universe model. The fact that combining cut-off physics effects, as described by the polymerization technique, together with modified gravity, here summarized by the emerging physical clock, makes regular and physically well-posed the quasi-classical Universe dynamics, leads us to a further implementation of this picture. On one hand, we could study more general cosmological models to check if the valuable features, here established, survive in the absence of strong space symmetry. On the other hand, we could search for more fundamental connections between modified gravity theories and quantum gravity proposals, as in \cite{Bombacigno:2019nua}.%
\begin{acknowledgments}
M.L.L. would like to thank Federico Manzoni and Corrado Marzano for their help during the development of this work. S.L.F. would like to thank the TAsP INFN initiative (Rome 1 section) for support.
\end{acknowledgments}
\bibliography{references}

%apsrev4-2.bst 2019-01-14 (MD) hand-edited version of apsrev4-1.bst
%Control: key (0)
%Control: author (8) initials jnrlst
%Control: editor formatted (1) identically to author
%Control: production of article title (0) allowed
%Control: page (0) single
%Control: year (1) truncated
%Control: production of eprint (0) enabled
\begin{thebibliography}{63}%
\makeatletter
\providecommand \@ifxundefined [1]{%
 \@ifx{#1\undefined}
}%
\providecommand \@ifnum [1]{%
 \ifnum #1\expandafter \@firstoftwo
 \else \expandafter \@secondoftwo
 \fi
}%
\providecommand \@ifx [1]{%
 \ifx #1\expandafter \@firstoftwo
 \else \expandafter \@secondoftwo
 \fi
}%
\providecommand \natexlab [1]{#1}%
\providecommand \enquote  [1]{``#1''}%
\providecommand \bibnamefont  [1]{#1}%
\providecommand \bibfnamefont [1]{#1}%
\providecommand \citenamefont [1]{#1}%
\providecommand \href@noop [0]{\@secondoftwo}%
\providecommand \href [0]{\begingroup \@sanitize@url \@href}%
\providecommand \@href[1]{\@@startlink{#1}\@@href}%
\providecommand \@@href[1]{\endgroup#1\@@endlink}%
\providecommand \@sanitize@url [0]{\catcode `\\12\catcode `\$12\catcode `\&12\catcode `\#12\catcode `\^12\catcode `\_12\catcode `\%12\relax}%
\providecommand \@@startlink[1]{}%
\providecommand \@@endlink[0]{}%
\providecommand \url  [0]{\begingroup\@sanitize@url \@url }%
\providecommand \@url [1]{\endgroup\@href {#1}{\urlprefix }}%
\providecommand \urlprefix  [0]{URL }%
\providecommand \Eprint [0]{\href }%
\providecommand \doibase [0]{https://doi.org/}%
\providecommand \selectlanguage [0]{\@gobble}%
\providecommand \bibinfo  [0]{\@secondoftwo}%
\providecommand \bibfield  [0]{\@secondoftwo}%
\providecommand \translation [1]{[#1]}%
\providecommand \BibitemOpen [0]{}%
\providecommand \bibitemStop [0]{}%
\providecommand \bibitemNoStop [0]{.\EOS\space}%
\providecommand \EOS [0]{\spacefactor3000\relax}%
\providecommand \BibitemShut  [1]{\csname bibitem#1\endcsname}%
\let\auto@bib@innerbib\@empty
%</preamble>
\bibitem [{\citenamefont {Misner}\ \emph {et~al.}(1973)\citenamefont {Misner}, \citenamefont {Thorne},\ and\ \citenamefont {Wheeler}}]{Misner:1973prb}%
  \BibitemOpen
  \bibfield  {author} {\bibinfo {author} {\bibfnamefont {C.~W.}\ \bibnamefont {Misner}}, \bibinfo {author} {\bibfnamefont {K.~S.}\ \bibnamefont {Thorne}},\ and\ \bibinfo {author} {\bibfnamefont {J.~A.}\ \bibnamefont {Wheeler}},\ }\href@noop {} {\emph {\bibinfo {title} {{Gravitation}}}}\ (\bibinfo  {publisher} {W. H. Freeman},\ \bibinfo {address} {San Francisco},\ \bibinfo {year} {1973})\BibitemShut {NoStop}%
\bibitem [{\citenamefont {Carter}(2009)}]{Carter:2009nex}%
  \BibitemOpen
  \bibfield  {author} {\bibinfo {author} {\bibfnamefont {B.}~\bibnamefont {Carter}},\ }\bibfield  {title} {\bibinfo {title} {{Republication of: Black hole equilibrium states}},\ }\href {https://doi.org/10.1007/s10714-009-0888-5} {\bibfield  {journal} {\bibinfo  {journal} {Gen. Rel. Grav.}\ }\textbf {\bibinfo {volume} {41}},\ \bibinfo {pages} {2873} (\bibinfo {year} {2009})}\BibitemShut {NoStop}%
\bibitem [{\citenamefont {Montani}\ \emph {et~al.}(2009)\citenamefont {Montani}, \citenamefont {Battisti}, \citenamefont {Benini},\ and\ \citenamefont {Imponente}}]{Montani:PrimCos}%
  \BibitemOpen
  \bibfield  {author} {\bibinfo {author} {\bibfnamefont {G.}~\bibnamefont {Montani}}, \bibinfo {author} {\bibfnamefont {M.~V.}\ \bibnamefont {Battisti}}, \bibinfo {author} {\bibfnamefont {R.}~\bibnamefont {Benini}},\ and\ \bibinfo {author} {\bibfnamefont {G.}~\bibnamefont {Imponente}},\ }\href@noop {} {\emph {\bibinfo {title} {Primordial cosmology}}}\ (\bibinfo  {publisher} {World Scientific},\ \bibinfo {address} {Singapore},\ \bibinfo {year} {2009})\BibitemShut {NoStop}%
\bibitem [{\citenamefont {Weinberg}(2008)}]{Weinberg:2008zzc}%
  \BibitemOpen
  \bibfield  {author} {\bibinfo {author} {\bibfnamefont {S.}~\bibnamefont {Weinberg}},\ }\href@noop {} {\emph {\bibinfo {title} {Cosmology}}}\ (\bibinfo {year} {2008})\BibitemShut {NoStop}%
\bibitem [{\citenamefont {Hawking}\ and\ \citenamefont {Ellis}(2023)}]{Hawking:1973uf}%
  \BibitemOpen
  \bibfield  {author} {\bibinfo {author} {\bibfnamefont {S.~W.}\ \bibnamefont {Hawking}}\ and\ \bibinfo {author} {\bibfnamefont {G.~F.~R.}\ \bibnamefont {Ellis}},\ }\href {https://doi.org/10.1017/9781009253161} {\emph {\bibinfo {title} {The Large Scale Structure of Space-Time}}},\ Cambridge Monographs on Mathematical Physics\ (\bibinfo  {publisher} {Cambridge University Press},\ \bibinfo {address} {UK},\ \bibinfo {year} {2023})\BibitemShut {NoStop}%
\bibitem [{\citenamefont {Belinskiĭ}\ \emph {et~al.}(1971)\citenamefont {Belinskiĭ}, \citenamefont {Lifshitz},\ and\ \citenamefont {Khalatnikov}}]{BKL_1971}%
  \BibitemOpen
  \bibfield  {author} {\bibinfo {author} {\bibfnamefont {V.~A.}\ \bibnamefont {Belinskiĭ}}, \bibinfo {author} {\bibfnamefont {E.~M.}\ \bibnamefont {Lifshitz}},\ and\ \bibinfo {author} {\bibfnamefont {I.~M.}\ \bibnamefont {Khalatnikov}},\ }\bibfield  {title} {\bibinfo {title} {Oscillatory approach to the singular point in relativistic cosmology},\ }\href {https://doi.org/10.1070/PU1971v013n06ABEH004279} {\bibfield  {journal} {\bibinfo  {journal} {Soviet Physics Uspekhi}\ }\textbf {\bibinfo {volume} {13}},\ \bibinfo {pages} {745} (\bibinfo {year} {1971})}\BibitemShut {NoStop}%
\bibitem [{\citenamefont {Belinsky}\ \emph {et~al.}(1982)\citenamefont {Belinsky}, \citenamefont {Khalatnikov},\ and\ \citenamefont {Lifshitz}}]{BKL_1982}%
  \BibitemOpen
  \bibfield  {author} {\bibinfo {author} {\bibfnamefont {V.~M.}\ \bibnamefont {Belinsky}}, \bibinfo {author} {\bibfnamefont {I.~M.}\ \bibnamefont {Khalatnikov}},\ and\ \bibinfo {author} {\bibfnamefont {E.~M.}\ \bibnamefont {Lifshitz}},\ }\bibfield  {title} {\bibinfo {title} {{A General Solution of the Einstein Equations with a Time Singularity}},\ }\href {https://doi.org/10.1080/00018738200101428} {\bibfield  {journal} {\bibinfo  {journal} {Adv. Phys.}\ }\textbf {\bibinfo {volume} {31}},\ \bibinfo {pages} {639} (\bibinfo {year} {1982})}\BibitemShut {NoStop}%
\bibitem [{\citenamefont {Kirillov}(1993)}]{kirillov_93}%
  \BibitemOpen
  \bibfield  {author} {\bibinfo {author} {\bibfnamefont {A.}~\bibnamefont {Kirillov}},\ }\bibfield  {title} {\bibinfo {title} {On the nature of the spatial distribution of metric inhomogeneities in the general solution of the einstein equations near a cosmological singularity},\ }\href@noop {} {\bibfield  {journal} {\bibinfo  {journal} {Sov. Phys. JETP}\ }\textbf {\bibinfo {volume} {76}} (\bibinfo {year} {1993})}\BibitemShut {NoStop}%
\bibitem [{\citenamefont {Montani}(1995)}]{Montani_1995}%
  \BibitemOpen
  \bibfield  {author} {\bibinfo {author} {\bibfnamefont {G.}~\bibnamefont {Montani}},\ }\bibfield  {title} {\bibinfo {title} {On the general behaviour of the universe near the cosmological singularity},\ }\href {https://doi.org/10.1088/0264-9381/12/10/010} {\bibfield  {journal} {\bibinfo  {journal} {Classical and Quantum Gravity}\ }\textbf {\bibinfo {volume} {12}},\ \bibinfo {pages} {2505} (\bibinfo {year} {1995})}\BibitemShut {NoStop}%
\bibitem [{\citenamefont {Imponente}\ and\ \citenamefont {Montani}(2001)}]{montani_imponente_01}%
  \BibitemOpen
  \bibfield  {author} {\bibinfo {author} {\bibfnamefont {G.}~\bibnamefont {Imponente}}\ and\ \bibinfo {author} {\bibfnamefont {G.}~\bibnamefont {Montani}},\ }\bibfield  {title} {\bibinfo {title} {Covariance of the mixmaster chaoticity},\ }\href {https://doi.org/10.1103/PhysRevD.63.103501} {\bibfield  {journal} {\bibinfo  {journal} {Phys. Rev. D}\ }\textbf {\bibinfo {volume} {63}},\ \bibinfo {pages} {103501} (\bibinfo {year} {2001})}\BibitemShut {NoStop}%
\bibitem [{\citenamefont {Benini}\ and\ \citenamefont {Montani}(2004)}]{montani_benini_04}%
  \BibitemOpen
  \bibfield  {author} {\bibinfo {author} {\bibfnamefont {R.}~\bibnamefont {Benini}}\ and\ \bibinfo {author} {\bibfnamefont {G.}~\bibnamefont {Montani}},\ }\bibfield  {title} {\bibinfo {title} {Frame independence of the inhomogeneous mixmaster chaos via misner-chitr\'e-like variables},\ }\href {https://doi.org/10.1103/PhysRevD.70.103527} {\bibfield  {journal} {\bibinfo  {journal} {Phys. Rev. D}\ }\textbf {\bibinfo {volume} {70}},\ \bibinfo {pages} {103527} (\bibinfo {year} {2004})}\BibitemShut {NoStop}%
\bibitem [{\citenamefont {Montani}\ \emph {et~al.}(2008)\citenamefont {Montani}, \citenamefont {Battisti}, \citenamefont {Benini},\ and\ \citenamefont {Imponente}}]{MONTANI_2008}%
  \BibitemOpen
  \bibfield  {author} {\bibinfo {author} {\bibfnamefont {G.}~\bibnamefont {Montani}}, \bibinfo {author} {\bibfnamefont {M.~V.}\ \bibnamefont {Battisti}}, \bibinfo {author} {\bibfnamefont {R.}~\bibnamefont {Benini}},\ and\ \bibinfo {author} {\bibfnamefont {G.}~\bibnamefont {Imponente}},\ }\bibfield  {title} {\bibinfo {title} {Classical and quantum features of the mixmaster singularity},\ }\href {https://doi.org/10.1142/s0217751x08040275} {\bibfield  {journal} {\bibinfo  {journal} {International Journal of Modern Physics A}\ }\textbf {\bibinfo {volume} {23}},\ \bibinfo {pages} {2353–2503} (\bibinfo {year} {2008})}\BibitemShut {NoStop}%
\bibitem [{\citenamefont {{Lo Franco}}\ and\ \citenamefont {Montani}(2025)}]{LOFRANCO2025100463}%
  \BibitemOpen
  \bibfield  {author} {\bibinfo {author} {\bibfnamefont {S.}~\bibnamefont {{Lo Franco}}}\ and\ \bibinfo {author} {\bibfnamefont {G.}~\bibnamefont {Montani}},\ }\bibfield  {title} {\bibinfo {title} {Is the bkl map an intrinsic feature of the quantum mixmaster universe?},\ }\href {https://doi.org/https://doi.org/10.1016/j.jheap.2025.100463} {\bibfield  {journal} {\bibinfo  {journal} {Journal of High Energy Astrophysics}\ ,\ \bibinfo {pages} {100463}} (\bibinfo {year} {2025})}\BibitemShut {NoStop}%
\bibitem [{\citenamefont {Cianfrani}\ \emph {et~al.}(2014)\citenamefont {Cianfrani}, \citenamefont {Lecian}, \citenamefont {Lulli},\ and\ \citenamefont {Montani}}]{montani_cqg}%
  \BibitemOpen
  \bibfield  {author} {\bibinfo {author} {\bibfnamefont {F.}~\bibnamefont {Cianfrani}}, \bibinfo {author} {\bibfnamefont {O.~M.}\ \bibnamefont {Lecian}}, \bibinfo {author} {\bibfnamefont {M.}~\bibnamefont {Lulli}},\ and\ \bibinfo {author} {\bibfnamefont {G.}~\bibnamefont {Montani}},\ }\href {https://doi.org/10.1142/8957} {\emph {\bibinfo {title} {Canonical Quantum Gravity}}}\ (\bibinfo  {publisher} {World Scientific},\ \bibinfo {address} {Singapore},\ \bibinfo {year} {2014})\ \Eprint {https://arxiv.org/abs/https://www.worldscientific.com/doi/pdf/10.1142/8957} {https://www.worldscientific.com/doi/pdf/10.1142/8957} \BibitemShut {NoStop}%
\bibitem [{\citenamefont {Thiemann}(2007)}]{Thiemann:2007pyv}%
  \BibitemOpen
  \bibfield  {author} {\bibinfo {author} {\bibfnamefont {T.}~\bibnamefont {Thiemann}},\ }\href {https://doi.org/10.1017/CBO9780511755682} {\emph {\bibinfo {title} {Modern Canonical Quantum General Relativity}}},\ Cambridge Monographs on Mathematical Physics\ (\bibinfo  {publisher} {Cambridge University Press},\ \bibinfo {address} {UK},\ \bibinfo {year} {2007})\BibitemShut {NoStop}%
\bibitem [{\citenamefont {Sotiriou}\ and\ \citenamefont {Faraoni}(2010)}]{Sotiriou_2010}%
  \BibitemOpen
  \bibfield  {author} {\bibinfo {author} {\bibfnamefont {T.~P.}\ \bibnamefont {Sotiriou}}\ and\ \bibinfo {author} {\bibfnamefont {V.}~\bibnamefont {Faraoni}},\ }\bibfield  {title} {\bibinfo {title} {f(r) theories of gravity},\ }\href {https://doi.org/10.1103/revmodphys.82.451} {\bibfield  {journal} {\bibinfo  {journal} {Reviews of Modern Physics}\ }\textbf {\bibinfo {volume} {82}},\ \bibinfo {pages} {451–497} (\bibinfo {year} {2010})}\BibitemShut {NoStop}%
\bibitem [{\citenamefont {Capozziello}\ \emph {et~al.}(2010)\citenamefont {Capozziello}, \citenamefont {De~Laurentis},\ and\ \citenamefont {Faraoni}}]{Capozziello:2009nq}%
  \BibitemOpen
  \bibfield  {author} {\bibinfo {author} {\bibfnamefont {S.}~\bibnamefont {Capozziello}}, \bibinfo {author} {\bibfnamefont {M.}~\bibnamefont {De~Laurentis}},\ and\ \bibinfo {author} {\bibfnamefont {V.}~\bibnamefont {Faraoni}},\ }\bibfield  {title} {\bibinfo {title} {{A Bird's eye view of f(R)-gravity}},\ }\href {https://doi.org/10.2174/1874381101003020049} {\bibfield  {journal} {\bibinfo  {journal} {Open Astron. J.}\ }\textbf {\bibinfo {volume} {3}},\ \bibinfo {pages} {49} (\bibinfo {year} {2010})},\ \Eprint {https://arxiv.org/abs/0909.4672} {arXiv:0909.4672 [gr-qc]} \BibitemShut {NoStop}%
\bibitem [{\citenamefont {Nojiri}\ \emph {et~al.}(2017)\citenamefont {Nojiri}, \citenamefont {Odintsov},\ and\ \citenamefont {Oikonomou}}]{Nojiri:2017ncd}%
  \BibitemOpen
  \bibfield  {author} {\bibinfo {author} {\bibfnamefont {S.}~\bibnamefont {Nojiri}}, \bibinfo {author} {\bibfnamefont {S.~D.}\ \bibnamefont {Odintsov}},\ and\ \bibinfo {author} {\bibfnamefont {V.~K.}\ \bibnamefont {Oikonomou}},\ }\bibfield  {title} {\bibinfo {title} {{Modified Gravity Theories on a Nutshell: Inflation, Bounce and Late-time Evolution}},\ }\href {https://doi.org/10.1016/j.physrep.2017.06.001} {\bibfield  {journal} {\bibinfo  {journal} {Phys. Rept.}\ }\textbf {\bibinfo {volume} {692}},\ \bibinfo {pages} {1} (\bibinfo {year} {2017})},\ \Eprint {https://arxiv.org/abs/1705.11098} {arXiv:1705.11098 [gr-qc]} \BibitemShut {NoStop}%
\bibitem [{\citenamefont {Blyth}\ and\ \citenamefont {Isham}(1975)}]{Blyth:1975is}%
  \BibitemOpen
  \bibfield  {author} {\bibinfo {author} {\bibfnamefont {W.~F.}\ \bibnamefont {Blyth}}\ and\ \bibinfo {author} {\bibfnamefont {C.~J.}\ \bibnamefont {Isham}},\ }\bibfield  {title} {\bibinfo {title} {{Quantization of a Friedmann Universe Filled with a Scalar Field}},\ }\href {https://doi.org/10.1103/PhysRevD.11.768} {\bibfield  {journal} {\bibinfo  {journal} {Phys. Rev. D}\ }\textbf {\bibinfo {volume} {11}},\ \bibinfo {pages} {768} (\bibinfo {year} {1975})}\BibitemShut {NoStop}%
\bibitem [{\citenamefont {Kirillov}\ and\ \citenamefont {Montani}(1997)}]{kirillov_97}%
  \BibitemOpen
  \bibfield  {author} {\bibinfo {author} {\bibfnamefont {A.}~\bibnamefont {Kirillov}}\ and\ \bibinfo {author} {\bibfnamefont {G.}~\bibnamefont {Montani}},\ }\bibfield  {title} {\bibinfo {title} {Origin of a classical space in quantum inhomogeneous models},\ }\href {https://doi.org/10.1134/1.567553} {\bibfield  {journal} {\bibinfo  {journal} {JETP Letters}\ }\textbf {\bibinfo {volume} {66}},\ \bibinfo {pages} {475} (\bibinfo {year} {1997})}\BibitemShut {NoStop}%
\bibitem [{\citenamefont {Benini}\ and\ \citenamefont {Montani}(2006)}]{montani_benini2006inhomogeneous}%
  \BibitemOpen
  \bibfield  {author} {\bibinfo {author} {\bibfnamefont {R.}~\bibnamefont {Benini}}\ and\ \bibinfo {author} {\bibfnamefont {G.}~\bibnamefont {Montani}},\ }\bibfield  {title} {\bibinfo {title} {Inhomogeneous quantum mixmaster: From classical towards quantum mechanics},\ }\href@noop {} {\bibfield  {journal} {\bibinfo  {journal} {Classical and Quantum Gravity}\ }\textbf {\bibinfo {volume} {24}},\ \bibinfo {pages} {387} (\bibinfo {year} {2006})}\BibitemShut {NoStop}%
\bibitem [{\citenamefont {Giovannetti}\ and\ \citenamefont {Montani}(2022)}]{Giovannetti:2022qje}%
  \BibitemOpen
  \bibfield  {author} {\bibinfo {author} {\bibfnamefont {E.}~\bibnamefont {Giovannetti}}\ and\ \bibinfo {author} {\bibfnamefont {G.}~\bibnamefont {Montani}},\ }\bibfield  {title} {\bibinfo {title} {{Is Bianchi I a bouncing cosmology in the Wheeler-DeWitt picture?}},\ }\href {https://doi.org/10.1103/PhysRevD.106.044053} {\bibfield  {journal} {\bibinfo  {journal} {Phys. Rev. D}\ }\textbf {\bibinfo {volume} {106}},\ \bibinfo {pages} {044053} (\bibinfo {year} {2022})},\ \Eprint {https://arxiv.org/abs/2203.01062} {arXiv:2203.01062 [gr-qc]} \BibitemShut {NoStop}%
\bibitem [{\citenamefont {Giovannetti}\ \emph {et~al.}(2023)\citenamefont {Giovannetti}, \citenamefont {Maione},\ and\ \citenamefont {Montani}}]{giovannetti_maione}%
  \BibitemOpen
  \bibfield  {author} {\bibinfo {author} {\bibfnamefont {E.}~\bibnamefont {Giovannetti}}, \bibinfo {author} {\bibfnamefont {F.}~\bibnamefont {Maione}},\ and\ \bibinfo {author} {\bibfnamefont {G.}~\bibnamefont {Montani}},\ }\bibfield  {title} {\bibinfo {title} {Quantum big bounce of the isotropic universe using relational time},\ }\bibfield  {journal} {\bibinfo  {journal} {Universe}\ }\textbf {\bibinfo {volume} {9}},\ \href {https://doi.org/10.3390/universe9080373} {10.3390/universe9080373} (\bibinfo {year} {2023})\BibitemShut {NoStop}%
\bibitem [{\citenamefont {Lo~Franco}\ and\ \citenamefont {Montani}(2024)}]{LoFranco:2024nss}%
  \BibitemOpen
  \bibfield  {author} {\bibinfo {author} {\bibfnamefont {S.}~\bibnamefont {Lo~Franco}}\ and\ \bibinfo {author} {\bibfnamefont {G.}~\bibnamefont {Montani}},\ }\bibfield  {title} {\bibinfo {title} {{Quantum Big-Bounce as a phenomenology of RQM in the Mini-superspace}},\ }\href {https://doi.org/10.1016/j.physletb.2024.138983} {\bibfield  {journal} {\bibinfo  {journal} {Phys. Lett. B}\ }\textbf {\bibinfo {volume} {857}},\ \bibinfo {pages} {138983} (\bibinfo {year} {2024})},\ \Eprint {https://arxiv.org/abs/2404.02802} {arXiv:2404.02802 [gr-qc]} \BibitemShut {NoStop}%
\bibitem [{\citenamefont {Ashtekar}(1986)}]{Ashtekar:1986yd}%
  \BibitemOpen
  \bibfield  {author} {\bibinfo {author} {\bibfnamefont {A.}~\bibnamefont {Ashtekar}},\ }\bibfield  {title} {\bibinfo {title} {{New Variables for Classical and Quantum Gravity}},\ }\href {https://doi.org/10.1103/PhysRevLett.57.2244} {\bibfield  {journal} {\bibinfo  {journal} {Phys. Rev. Lett.}\ }\textbf {\bibinfo {volume} {57}},\ \bibinfo {pages} {2244} (\bibinfo {year} {1986})}\BibitemShut {NoStop}%
\bibitem [{\citenamefont {Barbero~G.}(1995)}]{BarberoG:1994eia}%
  \BibitemOpen
  \bibfield  {author} {\bibinfo {author} {\bibfnamefont {J.~F.}\ \bibnamefont {Barbero~G.}},\ }\bibfield  {title} {\bibinfo {title} {{Real Ashtekar variables for Lorentzian signature space times}},\ }\href {https://doi.org/10.1103/PhysRevD.51.5507} {\bibfield  {journal} {\bibinfo  {journal} {Phys. Rev. D}\ }\textbf {\bibinfo {volume} {51}},\ \bibinfo {pages} {5507} (\bibinfo {year} {1995})},\ \Eprint {https://arxiv.org/abs/gr-qc/9410014} {arXiv:gr-qc/9410014} \BibitemShut {NoStop}%
\bibitem [{\citenamefont {Immirzi}(1997)}]{Immirzi:1996di}%
  \BibitemOpen
  \bibfield  {author} {\bibinfo {author} {\bibfnamefont {G.}~\bibnamefont {Immirzi}},\ }\bibfield  {title} {\bibinfo {title} {{Real and complex connections for canonical gravity}},\ }\href {https://doi.org/10.1088/0264-9381/14/10/002} {\bibfield  {journal} {\bibinfo  {journal} {Class. Quant. Grav.}\ }\textbf {\bibinfo {volume} {14}},\ \bibinfo {pages} {L177} (\bibinfo {year} {1997})},\ \Eprint {https://arxiv.org/abs/gr-qc/9612030} {arXiv:gr-qc/9612030} \BibitemShut {NoStop}%
\bibitem [{\citenamefont {Ashtekar}\ and\ \citenamefont {Singh}(2011)}]{Ashtekar_2011}%
  \BibitemOpen
  \bibfield  {author} {\bibinfo {author} {\bibfnamefont {A.}~\bibnamefont {Ashtekar}}\ and\ \bibinfo {author} {\bibfnamefont {P.}~\bibnamefont {Singh}},\ }\bibfield  {title} {\bibinfo {title} {Loop quantum cosmology: a status report},\ }\href {https://doi.org/10.1088/0264-9381/28/21/213001} {\bibfield  {journal} {\bibinfo  {journal} {Classical and Quantum Gravity}\ }\textbf {\bibinfo {volume} {28}},\ \bibinfo {pages} {213001} (\bibinfo {year} {2011})}\BibitemShut {NoStop}%
\bibitem [{\citenamefont {Bojowald}(2019)}]{Bojowald_2019}%
  \BibitemOpen
  \bibfield  {author} {\bibinfo {author} {\bibfnamefont {M.}~\bibnamefont {Bojowald}},\ }\bibfield  {title} {\bibinfo {title} {Effective field theory of loop quantum cosmology},\ }\href {https://doi.org/10.3390/universe5020044} {\bibfield  {journal} {\bibinfo  {journal} {Universe}\ }\textbf {\bibinfo {volume} {5}},\ \bibinfo {pages} {44} (\bibinfo {year} {2019})}\BibitemShut {NoStop}%
\bibitem [{\citenamefont {Bojowald}(2001)}]{Bojowald:2000pk}%
  \BibitemOpen
  \bibfield  {author} {\bibinfo {author} {\bibfnamefont {M.}~\bibnamefont {Bojowald}},\ }\bibfield  {title} {\bibinfo {title} {{Loop quantum cosmology. 4. Discrete time evolution}},\ }\href {https://doi.org/10.1088/0264-9381/18/6/308} {\bibfield  {journal} {\bibinfo  {journal} {Class. Quant. Grav.}\ }\textbf {\bibinfo {volume} {18}},\ \bibinfo {pages} {1071} (\bibinfo {year} {2001})},\ \Eprint {https://arxiv.org/abs/gr-qc/0008053} {arXiv:gr-qc/0008053} \BibitemShut {NoStop}%
\bibitem [{\citenamefont {Cianfrani}\ and\ \citenamefont {Montani}(2009)}]{Cianfrani:2008zv}%
  \BibitemOpen
  \bibfield  {author} {\bibinfo {author} {\bibfnamefont {F.}~\bibnamefont {Cianfrani}}\ and\ \bibinfo {author} {\bibfnamefont {G.}~\bibnamefont {Montani}},\ }\bibfield  {title} {\bibinfo {title} {Towards loop quantum gravity without the time gauge},\ }\href {https://doi.org/10.1103/PhysRevLett.102.091301} {\bibfield  {journal} {\bibinfo  {journal} {Phys. Rev. Lett.}\ }\textbf {\bibinfo {volume} {102}},\ \bibinfo {pages} {091301} (\bibinfo {year} {2009})},\ \Eprint {https://arxiv.org/abs/0811.1916} {arXiv:0811.1916 [gr-qc]} \BibitemShut {NoStop}%
\bibitem [{\citenamefont {Cianfrani}\ and\ \citenamefont {Montani}(2012)}]{Cianfrani:2011wg}%
  \BibitemOpen
  \bibfield  {author} {\bibinfo {author} {\bibfnamefont {F.}~\bibnamefont {Cianfrani}}\ and\ \bibinfo {author} {\bibfnamefont {G.}~\bibnamefont {Montani}},\ }\bibfield  {title} {\bibinfo {title} {Implications of the gauge-fixing in loop quantum cosmology},\ }\href {https://doi.org/10.1103/PhysRevD.85.024027} {\bibfield  {journal} {\bibinfo  {journal} {Phys. Rev. D}\ }\textbf {\bibinfo {volume} {85}},\ \bibinfo {pages} {024027} (\bibinfo {year} {2012})},\ \Eprint {https://arxiv.org/abs/1104.4546} {arXiv:1104.4546 [gr-qc]} \BibitemShut {NoStop}%
\bibitem [{\citenamefont {Bruno}\ and\ \citenamefont {Montani}(2023{\natexlab{a}})}]{Bruno:2023aco}%
  \BibitemOpen
  \bibfield  {author} {\bibinfo {author} {\bibfnamefont {M.}~\bibnamefont {Bruno}}\ and\ \bibinfo {author} {\bibfnamefont {G.}~\bibnamefont {Montani}},\ }\bibfield  {title} {\bibinfo {title} {{Is the diagonal case a general picture for loop quantum cosmology?}},\ }\href {https://doi.org/10.1103/PhysRevD.108.046003} {\bibfield  {journal} {\bibinfo  {journal} {Phys. Rev. D}\ }\textbf {\bibinfo {volume} {108}},\ \bibinfo {pages} {046003} (\bibinfo {year} {2023}{\natexlab{a}})},\ \Eprint {https://arxiv.org/abs/2306.10934} {arXiv:2306.10934 [gr-qc]} \BibitemShut {NoStop}%
\bibitem [{\citenamefont {Bruno}\ and\ \citenamefont {Montani}(2023{\natexlab{b}})}]{Bruno:2023all}%
  \BibitemOpen
  \bibfield  {author} {\bibinfo {author} {\bibfnamefont {M.}~\bibnamefont {Bruno}}\ and\ \bibinfo {author} {\bibfnamefont {G.}~\bibnamefont {Montani}},\ }\bibfield  {title} {\bibinfo {title} {{Loop quantum cosmology of nondiagonal Bianchi models}},\ }\href {https://doi.org/10.1103/PhysRevD.107.126013} {\bibfield  {journal} {\bibinfo  {journal} {Phys. Rev. D}\ }\textbf {\bibinfo {volume} {107}},\ \bibinfo {pages} {126013} (\bibinfo {year} {2023}{\natexlab{b}})},\ \Eprint {https://arxiv.org/abs/2302.03638} {arXiv:2302.03638 [gr-qc]} \BibitemShut {NoStop}%
\bibitem [{\citenamefont {Alesci}\ and\ \citenamefont {Cianfrani}(2013)}]{Alesci:2013xd}%
  \BibitemOpen
  \bibfield  {author} {\bibinfo {author} {\bibfnamefont {E.}~\bibnamefont {Alesci}}\ and\ \bibinfo {author} {\bibfnamefont {F.}~\bibnamefont {Cianfrani}},\ }\bibfield  {title} {\bibinfo {title} {{Quantum-Reduced Loop Gravity: Cosmology}},\ }\href {https://doi.org/10.1103/PhysRevD.87.083521} {\bibfield  {journal} {\bibinfo  {journal} {Phys. Rev. D}\ }\textbf {\bibinfo {volume} {87}},\ \bibinfo {pages} {083521} (\bibinfo {year} {2013})},\ \Eprint {https://arxiv.org/abs/1301.2245} {arXiv:1301.2245 [gr-qc]} \BibitemShut {NoStop}%
\bibitem [{\citenamefont {Alesci}\ and\ \citenamefont {Cianfrani}(2015)}]{Alesci:2014rra}%
  \BibitemOpen
  \bibfield  {author} {\bibinfo {author} {\bibfnamefont {E.}~\bibnamefont {Alesci}}\ and\ \bibinfo {author} {\bibfnamefont {F.}~\bibnamefont {Cianfrani}},\ }\bibfield  {title} {\bibinfo {title} {{Loop quantum cosmology from quantum reduced loop gravity}},\ }\href {https://doi.org/10.1209/0295-5075/111/40002} {\bibfield  {journal} {\bibinfo  {journal} {EPL}\ }\textbf {\bibinfo {volume} {111}},\ \bibinfo {pages} {40002} (\bibinfo {year} {2015})},\ \Eprint {https://arxiv.org/abs/1410.4788} {arXiv:1410.4788 [gr-qc]} \BibitemShut {NoStop}%
\bibitem [{\citenamefont {Alesci}\ and\ \citenamefont {Cianfrani}(2014)}]{Alesci:2014uha}%
  \BibitemOpen
  \bibfield  {author} {\bibinfo {author} {\bibfnamefont {E.}~\bibnamefont {Alesci}}\ and\ \bibinfo {author} {\bibfnamefont {F.}~\bibnamefont {Cianfrani}},\ }\bibfield  {title} {\bibinfo {title} {{Quantum Reduced Loop Gravity: Semiclassical limit}},\ }\href {https://doi.org/10.1103/PhysRevD.90.024006} {\bibfield  {journal} {\bibinfo  {journal} {Phys. Rev. D}\ }\textbf {\bibinfo {volume} {90}},\ \bibinfo {pages} {024006} (\bibinfo {year} {2014})},\ \Eprint {https://arxiv.org/abs/1402.3155} {arXiv:1402.3155 [gr-qc]} \BibitemShut {NoStop}%
\bibitem [{\citenamefont {Bojowald}(2002)}]{Bojowald:2002gz}%
  \BibitemOpen
  \bibfield  {author} {\bibinfo {author} {\bibfnamefont {M.}~\bibnamefont {Bojowald}},\ }\bibfield  {title} {\bibinfo {title} {{Isotropic loop quantum cosmology}},\ }\href {https://doi.org/10.1088/0264-9381/19/10/313} {\bibfield  {journal} {\bibinfo  {journal} {Class. Quant. Grav.}\ }\textbf {\bibinfo {volume} {19}},\ \bibinfo {pages} {2717} (\bibinfo {year} {2002})},\ \Eprint {https://arxiv.org/abs/gr-qc/0202077} {arXiv:gr-qc/0202077} \BibitemShut {NoStop}%
\bibitem [{\citenamefont {Bojowald}(2004)}]{Bojowald:2004af}%
  \BibitemOpen
  \bibfield  {author} {\bibinfo {author} {\bibfnamefont {M.}~\bibnamefont {Bojowald}},\ }\bibfield  {title} {\bibinfo {title} {{Spherically symmetric quantum geometry: States and basic operators}},\ }\href {https://doi.org/10.1088/0264-9381/21/15/008} {\bibfield  {journal} {\bibinfo  {journal} {Class. Quant. Grav.}\ }\textbf {\bibinfo {volume} {21}},\ \bibinfo {pages} {3733} (\bibinfo {year} {2004})},\ \Eprint {https://arxiv.org/abs/gr-qc/0407017} {arXiv:gr-qc/0407017} \BibitemShut {NoStop}%
\bibitem [{\citenamefont {Ashtekar}\ and\ \citenamefont {Bojowald}(2006)}]{Ashtekar:2005qt}%
  \BibitemOpen
  \bibfield  {author} {\bibinfo {author} {\bibfnamefont {A.}~\bibnamefont {Ashtekar}}\ and\ \bibinfo {author} {\bibfnamefont {M.}~\bibnamefont {Bojowald}},\ }\bibfield  {title} {\bibinfo {title} {{Quantum geometry and the Schwarzschild singularity}},\ }\href {https://doi.org/10.1088/0264-9381/23/2/008} {\bibfield  {journal} {\bibinfo  {journal} {Class. Quant. Grav.}\ }\textbf {\bibinfo {volume} {23}},\ \bibinfo {pages} {391} (\bibinfo {year} {2006})},\ \Eprint {https://arxiv.org/abs/gr-qc/0509075} {arXiv:gr-qc/0509075} \BibitemShut {NoStop}%
\bibitem [{\citenamefont {Ashtekar}\ \emph {et~al.}(2006)\citenamefont {Ashtekar}, \citenamefont {Pawlowski},\ and\ \citenamefont {Singh}}]{Ashtekar:2006wn}%
  \BibitemOpen
  \bibfield  {author} {\bibinfo {author} {\bibfnamefont {A.}~\bibnamefont {Ashtekar}}, \bibinfo {author} {\bibfnamefont {T.}~\bibnamefont {Pawlowski}},\ and\ \bibinfo {author} {\bibfnamefont {P.}~\bibnamefont {Singh}},\ }\bibfield  {title} {\bibinfo {title} {{Quantum Nature of the Big Bang: Improved dynamics}},\ }\href {https://doi.org/10.1103/PhysRevD.74.084003} {\bibfield  {journal} {\bibinfo  {journal} {Phys. Rev. D}\ }\textbf {\bibinfo {volume} {74}},\ \bibinfo {pages} {084003} (\bibinfo {year} {2006})},\ \Eprint {https://arxiv.org/abs/gr-qc/0607039} {arXiv:gr-qc/0607039} \BibitemShut {NoStop}%
\bibitem [{\citenamefont {Bombacigno}\ \emph {et~al.}(2016)\citenamefont {Bombacigno}, \citenamefont {Cianfrani},\ and\ \citenamefont {Montani}}]{Bombacigno:2016siz}%
  \BibitemOpen
  \bibfield  {author} {\bibinfo {author} {\bibfnamefont {F.}~\bibnamefont {Bombacigno}}, \bibinfo {author} {\bibfnamefont {F.}~\bibnamefont {Cianfrani}},\ and\ \bibinfo {author} {\bibfnamefont {G.}~\bibnamefont {Montani}},\ }\bibfield  {title} {\bibinfo {title} {{Big-Bounce cosmology in the presence of Immirzi field}},\ }\href {https://doi.org/10.1103/PhysRevD.94.064021} {\bibfield  {journal} {\bibinfo  {journal} {Phys. Rev. D}\ }\textbf {\bibinfo {volume} {94}},\ \bibinfo {pages} {064021} (\bibinfo {year} {2016})},\ \Eprint {https://arxiv.org/abs/1607.00910} {arXiv:1607.00910 [gr-qc]} \BibitemShut {NoStop}%
\bibitem [{\citenamefont {Bamba}\ \emph {et~al.}(2014)\citenamefont {Bamba}, \citenamefont {Makarenko}, \citenamefont {Myagky}, \citenamefont {Nojiri},\ and\ \citenamefont {Odintsov}}]{Bamba:2013fha}%
  \BibitemOpen
  \bibfield  {author} {\bibinfo {author} {\bibfnamefont {K.}~\bibnamefont {Bamba}}, \bibinfo {author} {\bibfnamefont {A.~N.}\ \bibnamefont {Makarenko}}, \bibinfo {author} {\bibfnamefont {A.~N.}\ \bibnamefont {Myagky}}, \bibinfo {author} {\bibfnamefont {S.}~\bibnamefont {Nojiri}},\ and\ \bibinfo {author} {\bibfnamefont {S.~D.}\ \bibnamefont {Odintsov}},\ }\bibfield  {title} {\bibinfo {title} {{Bounce cosmology from $F(R)$ gravity and $F(R)$ bigravity}},\ }\href {https://doi.org/10.1088/1475-7516/2014/01/008} {\bibfield  {journal} {\bibinfo  {journal} {JCAP}\ }\textbf {\bibinfo {volume} {01}},\ \bibinfo {pages} {008}},\ \Eprint {https://arxiv.org/abs/1309.3748} {arXiv:1309.3748 [hep-th]} \BibitemShut {NoStop}%
\bibitem [{\citenamefont {Odintsov}\ and\ \citenamefont {Oikonomou}(2014)}]{Odintsov:2014gea}%
  \BibitemOpen
  \bibfield  {author} {\bibinfo {author} {\bibfnamefont {S.~D.}\ \bibnamefont {Odintsov}}\ and\ \bibinfo {author} {\bibfnamefont {V.~K.}\ \bibnamefont {Oikonomou}},\ }\bibfield  {title} {\bibinfo {title} {{Matter Bounce Loop Quantum Cosmology from $F(R)$ Gravity}},\ }\href {https://doi.org/10.1103/PhysRevD.90.124083} {\bibfield  {journal} {\bibinfo  {journal} {Phys. Rev. D}\ }\textbf {\bibinfo {volume} {90}},\ \bibinfo {pages} {124083} (\bibinfo {year} {2014})},\ \Eprint {https://arxiv.org/abs/1410.8183} {arXiv:1410.8183 [gr-qc]} \BibitemShut {NoStop}%
\bibitem [{\citenamefont {Odintsov}\ \emph {et~al.}(2015)\citenamefont {Odintsov}, \citenamefont {Oikonomou},\ and\ \citenamefont {Saridakis}}]{Odintsov:2015uca}%
  \BibitemOpen
  \bibfield  {author} {\bibinfo {author} {\bibfnamefont {S.~D.}\ \bibnamefont {Odintsov}}, \bibinfo {author} {\bibfnamefont {V.~K.}\ \bibnamefont {Oikonomou}},\ and\ \bibinfo {author} {\bibfnamefont {E.~N.}\ \bibnamefont {Saridakis}},\ }\bibfield  {title} {\bibinfo {title} {{Superbounce and Loop Quantum Ekpyrotic Cosmologies from Modified Gravity: $F(R)$, $F(G)$ and $F(T)$ Theories}},\ }\href {https://doi.org/10.1016/j.aop.2015.08.021} {\bibfield  {journal} {\bibinfo  {journal} {Annals Phys.}\ }\textbf {\bibinfo {volume} {363}},\ \bibinfo {pages} {141} (\bibinfo {year} {2015})},\ \Eprint {https://arxiv.org/abs/1501.06591} {arXiv:1501.06591 [gr-qc]} \BibitemShut {NoStop}%
\bibitem [{\citenamefont {Bamonti}\ \emph {et~al.}(2022)\citenamefont {Bamonti}, \citenamefont {Costantini},\ and\ \citenamefont {Montani}}]{Bamonti:2021jmg}%
  \BibitemOpen
  \bibfield  {author} {\bibinfo {author} {\bibfnamefont {N.}~\bibnamefont {Bamonti}}, \bibinfo {author} {\bibfnamefont {A.}~\bibnamefont {Costantini}},\ and\ \bibinfo {author} {\bibfnamefont {G.}~\bibnamefont {Montani}},\ }\bibfield  {title} {\bibinfo {title} {{Features of the primordial Universe in f(R)-gravity as viewed in the Jordan frame}},\ }\href {https://doi.org/10.1088/1361-6382/ac7694} {\bibfield  {journal} {\bibinfo  {journal} {Class. Quant. Grav.}\ }\textbf {\bibinfo {volume} {39}},\ \bibinfo {pages} {175011} (\bibinfo {year} {2022})},\ \Eprint {https://arxiv.org/abs/2103.17063} {arXiv:2103.17063 [gr-qc]} \BibitemShut {NoStop}%
\bibitem [{\citenamefont {Barragan}\ \emph {et~al.}(2009)\citenamefont {Barragan}, \citenamefont {Olmo},\ and\ \citenamefont {Sanchis-Alepuz}}]{Barragan:2009sq}%
  \BibitemOpen
  \bibfield  {author} {\bibinfo {author} {\bibfnamefont {C.}~\bibnamefont {Barragan}}, \bibinfo {author} {\bibfnamefont {G.~J.}\ \bibnamefont {Olmo}},\ and\ \bibinfo {author} {\bibfnamefont {H.}~\bibnamefont {Sanchis-Alepuz}},\ }\bibfield  {title} {\bibinfo {title} {{Bouncing Cosmologies in Palatini f(R) Gravity}},\ }\href {https://doi.org/10.1103/PhysRevD.80.024016} {\bibfield  {journal} {\bibinfo  {journal} {Phys. Rev. D}\ }\textbf {\bibinfo {volume} {80}},\ \bibinfo {pages} {024016} (\bibinfo {year} {2009})},\ \Eprint {https://arxiv.org/abs/0907.0318} {arXiv:0907.0318 [gr-qc]} \BibitemShut {NoStop}%
\bibitem [{\citenamefont {Barragan}\ and\ \citenamefont {Olmo}(2010)}]{Barragan:2010qb}%
  \BibitemOpen
  \bibfield  {author} {\bibinfo {author} {\bibfnamefont {C.}~\bibnamefont {Barragan}}\ and\ \bibinfo {author} {\bibfnamefont {G.~J.}\ \bibnamefont {Olmo}},\ }\bibfield  {title} {\bibinfo {title} {{Isotropic and Anisotropic Bouncing Cosmologies in Palatini Gravity}},\ }\href {https://doi.org/10.1103/PhysRevD.82.084015} {\bibfield  {journal} {\bibinfo  {journal} {Phys. Rev. D}\ }\textbf {\bibinfo {volume} {82}},\ \bibinfo {pages} {084015} (\bibinfo {year} {2010})},\ \Eprint {https://arxiv.org/abs/1005.4136} {arXiv:1005.4136 [gr-qc]} \BibitemShut {NoStop}%
\bibitem [{\citenamefont {Bombacigno}\ \emph {et~al.}(2021{\natexlab{a}})\citenamefont {Bombacigno}, \citenamefont {Boudet}, \citenamefont {Olmo},\ and\ \citenamefont {Montani}}]{Bombacigno:2021bpk}%
  \BibitemOpen
  \bibfield  {author} {\bibinfo {author} {\bibfnamefont {F.}~\bibnamefont {Bombacigno}}, \bibinfo {author} {\bibfnamefont {S.}~\bibnamefont {Boudet}}, \bibinfo {author} {\bibfnamefont {G.~J.}\ \bibnamefont {Olmo}},\ and\ \bibinfo {author} {\bibfnamefont {G.}~\bibnamefont {Montani}},\ }\bibfield  {title} {\bibinfo {title} {{Big bounce and future time singularity resolution in Bianchi I cosmologies: The projective invariant Nieh-Yan case}},\ }\href {https://doi.org/10.1103/PhysRevD.103.124031} {\bibfield  {journal} {\bibinfo  {journal} {Phys. Rev. D}\ }\textbf {\bibinfo {volume} {103}},\ \bibinfo {pages} {124031} (\bibinfo {year} {2021}{\natexlab{a}})},\ \Eprint {https://arxiv.org/abs/2105.06870} {arXiv:2105.06870 [gr-qc]} \BibitemShut {NoStop}%
\bibitem [{\citenamefont {Cicoli}\ \emph {et~al.}(2024)\citenamefont {Cicoli}, \citenamefont {Conlon}, \citenamefont {Maharana}, \citenamefont {Parameswaran}, \citenamefont {Quevedo},\ and\ \citenamefont {Zavala}}]{CICOLI20241}%
  \BibitemOpen
  \bibfield  {author} {\bibinfo {author} {\bibfnamefont {M.}~\bibnamefont {Cicoli}}, \bibinfo {author} {\bibfnamefont {J.~P.}\ \bibnamefont {Conlon}}, \bibinfo {author} {\bibfnamefont {A.}~\bibnamefont {Maharana}}, \bibinfo {author} {\bibfnamefont {S.}~\bibnamefont {Parameswaran}}, \bibinfo {author} {\bibfnamefont {F.}~\bibnamefont {Quevedo}},\ and\ \bibinfo {author} {\bibfnamefont {I.}~\bibnamefont {Zavala}},\ }\bibfield  {title} {\bibinfo {title} {String cosmology: From the early universe to today},\ }\href {https://doi.org/https://doi.org/10.1016/j.physrep.2024.01.002} {\bibfield  {journal} {\bibinfo  {journal} {Physics Reports}\ }\textbf {\bibinfo {volume} {1059}},\ \bibinfo {pages} {1} (\bibinfo {year} {2024})},\ \bibinfo {note} {string Cosmology: from the Early Universe to Today}\BibitemShut {NoStop}%
\bibitem [{\citenamefont {Amor{\'o}s}\ \emph {et~al.}(2014)\citenamefont {Amor{\'o}s}, \citenamefont {de~Haro},\ and\ \citenamefont {Odintsov}}]{Amoros:2014tha}%
  \BibitemOpen
  \bibfield  {author} {\bibinfo {author} {\bibfnamefont {J.}~\bibnamefont {Amor{\'o}s}}, \bibinfo {author} {\bibfnamefont {J.}~\bibnamefont {de~Haro}},\ and\ \bibinfo {author} {\bibfnamefont {S.~D.}\ \bibnamefont {Odintsov}},\ }\bibfield  {title} {\bibinfo {title} {{$R+\alpha R^2$ Loop Quantum Cosmology}},\ }\href {https://doi.org/10.1103/PhysRevD.89.104010} {\bibfield  {journal} {\bibinfo  {journal} {Phys. Rev. D}\ }\textbf {\bibinfo {volume} {89}},\ \bibinfo {pages} {104010} (\bibinfo {year} {2014})},\ \Eprint {https://arxiv.org/abs/1402.3071} {arXiv:1402.3071 [gr-qc]} \BibitemShut {NoStop}%
\bibitem [{\citenamefont {De~Angelis}\ \emph {et~al.}(2021)\citenamefont {De~Angelis}, \citenamefont {Figurato},\ and\ \citenamefont {Montani}}]{DeAngelis:2021afq}%
  \BibitemOpen
  \bibfield  {author} {\bibinfo {author} {\bibfnamefont {M.}~\bibnamefont {De~Angelis}}, \bibinfo {author} {\bibfnamefont {L.}~\bibnamefont {Figurato}},\ and\ \bibinfo {author} {\bibfnamefont {G.}~\bibnamefont {Montani}},\ }\bibfield  {title} {\bibinfo {title} {{Quantum dynamics of the isotropic universe in metric f(R) gravity}},\ }\href {https://doi.org/10.1103/PhysRevD.104.024054} {\bibfield  {journal} {\bibinfo  {journal} {Phys. Rev. D}\ }\textbf {\bibinfo {volume} {104}},\ \bibinfo {pages} {024054} (\bibinfo {year} {2021})},\ \Eprint {https://arxiv.org/abs/2105.02934} {arXiv:2105.02934 [gr-qc]} \BibitemShut {NoStop}%
\bibitem [{\citenamefont {De~Angelis}\ and\ \citenamefont {Montani}(2023)}]{DeAngelis:2022qhm}%
  \BibitemOpen
  \bibfield  {author} {\bibinfo {author} {\bibfnamefont {M.}~\bibnamefont {De~Angelis}}\ and\ \bibinfo {author} {\bibfnamefont {G.}~\bibnamefont {Montani}},\ }\bibfield  {title} {\bibinfo {title} {{On the emergence of a classical isotropic universe from a quantum f(R) Bianchi cosmology in the Jordan frame}},\ }\href {https://doi.org/10.1140/epjc/s10052-023-11454-6} {\bibfield  {journal} {\bibinfo  {journal} {Eur. Phys. J. C}\ }\textbf {\bibinfo {volume} {83}},\ \bibinfo {pages} {285} (\bibinfo {year} {2023})},\ \Eprint {https://arxiv.org/abs/2207.14683} {arXiv:2207.14683 [gr-qc]} \BibitemShut {NoStop}%
\bibitem [{\citenamefont {Corichi}\ \emph {et~al.}(2007)\citenamefont {Corichi}, \citenamefont {Vukasinac},\ and\ \citenamefont {Zapata}}]{Corichi:2007tf}%
  \BibitemOpen
  \bibfield  {author} {\bibinfo {author} {\bibfnamefont {A.}~\bibnamefont {Corichi}}, \bibinfo {author} {\bibfnamefont {T.}~\bibnamefont {Vukasinac}},\ and\ \bibinfo {author} {\bibfnamefont {J.~A.}\ \bibnamefont {Zapata}},\ }\bibfield  {title} {\bibinfo {title} {{Polymer Quantum Mechanics and its Continuum Limit}},\ }\href {https://doi.org/10.1103/PhysRevD.76.044016} {\bibfield  {journal} {\bibinfo  {journal} {Phys. Rev. D}\ }\textbf {\bibinfo {volume} {76}},\ \bibinfo {pages} {044016} (\bibinfo {year} {2007})},\ \Eprint {https://arxiv.org/abs/0704.0007} {arXiv:0704.0007 [gr-qc]} \BibitemShut {NoStop}%
\bibitem [{\citenamefont {Barca}\ \emph {et~al.}(2021)\citenamefont {Barca}, \citenamefont {Giovannetti},\ and\ \citenamefont {Montani}}]{Barca:2021qdn}%
  \BibitemOpen
  \bibfield  {author} {\bibinfo {author} {\bibfnamefont {G.}~\bibnamefont {Barca}}, \bibinfo {author} {\bibfnamefont {E.}~\bibnamefont {Giovannetti}},\ and\ \bibinfo {author} {\bibfnamefont {G.}~\bibnamefont {Montani}},\ }\bibfield  {title} {\bibinfo {title} {{An Overview on the Nature of the Bounce in LQC and PQM}},\ }\href {https://doi.org/10.3390/universe7090327} {\bibfield  {journal} {\bibinfo  {journal} {Universe}\ }\textbf {\bibinfo {volume} {7}},\ \bibinfo {pages} {327} (\bibinfo {year} {2021})},\ \Eprint {https://arxiv.org/abs/2109.08645} {arXiv:2109.08645 [gr-qc]} \BibitemShut {NoStop}%
\bibitem [{\citenamefont {Montani}\ \emph {et~al.}(2019)\citenamefont {Montani}, \citenamefont {Mantero}, \citenamefont {Bombacigno}, \citenamefont {Cianfrani},\ and\ \citenamefont {Barca}}]{Montani:2018uay}%
  \BibitemOpen
  \bibfield  {author} {\bibinfo {author} {\bibfnamefont {G.}~\bibnamefont {Montani}}, \bibinfo {author} {\bibfnamefont {C.}~\bibnamefont {Mantero}}, \bibinfo {author} {\bibfnamefont {F.}~\bibnamefont {Bombacigno}}, \bibinfo {author} {\bibfnamefont {F.}~\bibnamefont {Cianfrani}},\ and\ \bibinfo {author} {\bibfnamefont {G.}~\bibnamefont {Barca}},\ }\bibfield  {title} {\bibinfo {title} {{Semiclassical and quantum analysis of the isotropic Universe in the polymer paradigm}},\ }\href {https://doi.org/10.1103/PhysRevD.99.063534} {\bibfield  {journal} {\bibinfo  {journal} {Phys. Rev. D}\ }\textbf {\bibinfo {volume} {99}},\ \bibinfo {pages} {063534} (\bibinfo {year} {2019})},\ \Eprint {https://arxiv.org/abs/1806.10364} {arXiv:1806.10364 [gr-qc]} \BibitemShut {NoStop}%
\bibitem [{\citenamefont {Giovannetti}\ \emph {et~al.}(2022)\citenamefont {Giovannetti}, \citenamefont {Barca}, \citenamefont {Mandini},\ and\ \citenamefont {Montani}}]{Giovannetti:2020nte}%
  \BibitemOpen
  \bibfield  {author} {\bibinfo {author} {\bibfnamefont {E.}~\bibnamefont {Giovannetti}}, \bibinfo {author} {\bibfnamefont {G.}~\bibnamefont {Barca}}, \bibinfo {author} {\bibfnamefont {F.}~\bibnamefont {Mandini}},\ and\ \bibinfo {author} {\bibfnamefont {G.}~\bibnamefont {Montani}},\ }\bibfield  {title} {\bibinfo {title} {{Polymer Dynamics of Isotropic Universe in Ashtekar and in Volume Variables}},\ }\href {https://doi.org/10.3390/universe8060302} {\bibfield  {journal} {\bibinfo  {journal} {Universe}\ }\textbf {\bibinfo {volume} {8}},\ \bibinfo {pages} {302} (\bibinfo {year} {2022})},\ \Eprint {https://arxiv.org/abs/2006.10614} {arXiv:2006.10614 [gr-qc]} \BibitemShut {NoStop}%
\bibitem [{\citenamefont {Moretti}\ \emph {et~al.}(2019)\citenamefont {Moretti}, \citenamefont {Bombacigno},\ and\ \citenamefont {Montani}}]{Moretti:2019yhs}%
  \BibitemOpen
  \bibfield  {author} {\bibinfo {author} {\bibfnamefont {F.}~\bibnamefont {Moretti}}, \bibinfo {author} {\bibfnamefont {F.}~\bibnamefont {Bombacigno}},\ and\ \bibinfo {author} {\bibfnamefont {G.}~\bibnamefont {Montani}},\ }\bibfield  {title} {\bibinfo {title} {{Gauge invariant formulation of metric $f(R)$ gravity for gravitational waves}},\ }\href {https://doi.org/10.1103/PhysRevD.100.084014} {\bibfield  {journal} {\bibinfo  {journal} {Phys. Rev. D}\ }\textbf {\bibinfo {volume} {100}},\ \bibinfo {pages} {084014} (\bibinfo {year} {2019})},\ \Eprint {https://arxiv.org/abs/1906.01899} {arXiv:1906.01899 [gr-qc]} \BibitemShut {NoStop}%
\bibitem [{\citenamefont {Ashtekar}\ and\ \citenamefont {Gupt}(2015)}]{Ashtekar_2015_effective}%
  \BibitemOpen
  \bibfield  {author} {\bibinfo {author} {\bibfnamefont {A.}~\bibnamefont {Ashtekar}}\ and\ \bibinfo {author} {\bibfnamefont {B.}~\bibnamefont {Gupt}},\ }\bibfield  {title} {\bibinfo {title} {Generalized effective description of loop quantum cosmology},\ }\bibfield  {journal} {\bibinfo  {journal} {Physical Review D}\ }\textbf {\bibinfo {volume} {92}},\ \href {https://doi.org/10.1103/physrevd.92.084060} {10.1103/physrevd.92.084060} (\bibinfo {year} {2015})\BibitemShut {NoStop}%
\bibitem [{\citenamefont {Singh}\ and\ \citenamefont {Vandersloot}(2005)}]{singh_2005_effective}%
  \BibitemOpen
  \bibfield  {author} {\bibinfo {author} {\bibfnamefont {P.}~\bibnamefont {Singh}}\ and\ \bibinfo {author} {\bibfnamefont {K.}~\bibnamefont {Vandersloot}},\ }\bibfield  {title} {\bibinfo {title} {Semiclassical states, effective dynamics, and classical emergence in loop quantum cosmology},\ }\href {https://doi.org/10.1103/PhysRevD.72.084004} {\bibfield  {journal} {\bibinfo  {journal} {Phys. Rev. D}\ }\textbf {\bibinfo {volume} {72}},\ \bibinfo {pages} {084004} (\bibinfo {year} {2005})}\BibitemShut {NoStop}%
\bibitem [{\citenamefont {Bombacigno}\ \emph {et~al.}(2021{\natexlab{b}})\citenamefont {Bombacigno}, \citenamefont {Boudet},\ and\ \citenamefont {Montani}}]{Bombacigno:2019nua}%
  \BibitemOpen
  \bibfield  {author} {\bibinfo {author} {\bibfnamefont {F.}~\bibnamefont {Bombacigno}}, \bibinfo {author} {\bibfnamefont {S.}~\bibnamefont {Boudet}},\ and\ \bibinfo {author} {\bibfnamefont {G.}~\bibnamefont {Montani}},\ }\bibfield  {title} {\bibinfo {title} {{Generalized Ashtekar variables for Palatini $f(\mathcal {R})$ models}},\ }\href {https://doi.org/10.1016/j.nuclphysb.2020.115281} {\bibfield  {journal} {\bibinfo  {journal} {Nucl. Phys. B}\ }\textbf {\bibinfo {volume} {963}},\ \bibinfo {pages} {115281} (\bibinfo {year} {2021}{\natexlab{b}})},\ \Eprint {https://arxiv.org/abs/1911.09066} {arXiv:1911.09066 [gr-qc]} \BibitemShut {NoStop}%
\bibitem [{\citenamefont {Kirillov}\ and\ \citenamefont {Montani}(2002)}]{Kirillov:2002kc}%
  \BibitemOpen
  \bibfield  {author} {\bibinfo {author} {\bibfnamefont {A.~A.}\ \bibnamefont {Kirillov}}\ and\ \bibinfo {author} {\bibfnamefont {G.}~\bibnamefont {Montani}},\ }\bibfield  {title} {\bibinfo {title} {{Quasiisotropization of the inhomogeneous mixmaster universe induced by an inflationary process}},\ }\href {https://doi.org/10.1103/PhysRevD.66.064010} {\bibfield  {journal} {\bibinfo  {journal} {Phys. Rev. D}\ }\textbf {\bibinfo {volume} {66}},\ \bibinfo {pages} {064010} (\bibinfo {year} {2002})},\ \Eprint {https://arxiv.org/abs/gr-qc/0209054} {arXiv:gr-qc/0209054} \BibitemShut {NoStop}%
\bibitem [{\citenamefont {Hall}()}]{Hall2013}%
  \BibitemOpen
  \bibfield  {author} {\bibinfo {author} {\bibfnamefont {B.~C.}\ \bibnamefont {Hall}},\ }\href@noop {} {\emph {\bibinfo {title} {Quantum Theory for Mathematicians}}},\ \bibinfo {series} {Graduate Texts in Mathematics}\ No.\ \bibinfo {number} {267}\ (\bibinfo  {publisher} {Springer New York})\BibitemShut {NoStop}%
\end{thebibliography}%
\end{document}